\documentclass[11pt]{paper}
\usepackage{lmodern}

\usepackage[T1]{fontenc}
\usepackage[latin9]{inputenc}
\usepackage[letterpaper]{geometry}
\usepackage{array}
\usepackage{prettyref}
\usepackage{amsthm}
\usepackage{amsmath}
\usepackage{amssymb}
\usepackage{graphicx}

\makeatletter

\providecommand{\tabularnewline}{\\}

\numberwithin{equation}{section}
\numberwithin{figure}{section}
\newcommand{\lyxaddress}[1]{
\par {\raggedright #1
\vspace{1.4em}
\noindent\par}
}
\theoremstyle{plain}
\newtheorem{thm}{\protect\theoremname}

\usepackage{algpseudocode}
\newcommand {\norm} [1] { \lVert #1 \rVert}

\newcommand {\abs} [1] {\left| #1 \right|}

\makeatother

\providecommand{\theoremname}{Theorem}

\begin{document}

\title{Metric Entropy and the Optimal Prediction of Chaotic Signals}

\author{Divakar Viswanath, Xuan Liang, and Kirill Serkh}

\maketitle

\lyxaddress{divakar@umich.edu, x5liang@ucsd.edu, kserkh@umich.edu. Address: (DV
and KS) Department of Mathematics, University of Michigan, Ann Arbor;
(XL) Department of Economics, University of California at San Diego.}
\begin{abstract}
Suppose we are given a time series or a signal $x(t)$ for $0\leq t\leq T$.
We consider the problem of predicting the signal in the interval $T<t\leq T+t_{f}$
from a knowledge of its history and nothing more. We ask the following
question: what is the largest value of $t_{f}$ for which a prediction
can be made? We show that the answer to this question is contained
in a fundamental result of information theory due to Wyner, Ziv, Ornstein,
and Weiss (1989, 1992). In particular, for the class of chaotic signals,
the upper bound is $t_{f}\leq\log_{2}T/H$ in the limit $T\rightarrow\infty$,
with $H$ being entropy in a sense that is explained in the text.

If $\bigl|x(T-s)-x(t^{\ast}-s)\bigr|$ is small for $0\leq s\leq\tau$,
where $\tau$ is of the order of a characteristic time scale, the
pattern of events leading up to $t=T$ is similar to the pattern of
events leading up to $t=t^{\ast}$. It is reasonable to expect $x(t^{\ast}+t_{f})$
to be a good predictor of $x(T+t_{f}).$ All existing methods for
prediction use this idea in one way or another. Unfortunately, this
intuitively reasonable idea is fundamentally deficient and all existing
methods fall well short of the Wyner-Ziv entropy bound on $t_{f}$.
An optimal predictor should decompose the distance between the pattern
of events leading up to $t=T$ and the pattern leading up to $t=t^{\ast}$
into stable and unstable components. A good match should have suitably
small unstable components but will in general allow stable components
which are as large as the tolerance for correct prediction. For the
special case of toral automorphisms, we use Padé approximants and
derive a predictor which has these properties and which seems to point
the way to the derivation of a more general optimal predictor.
\end{abstract}

\section{Introduction}

We consider the problem of predicting a signal or a time series $x(t)$
in the time interval $T<t\leq T+t_{f}$ assuming knowledge of the
signal in the time interval $0\leq t\leq T$. The signal is assumed
to originate from a deterministic dynamical system but we assume no
knowledge of the physical model. We consider the signal to be known
but assume no knowledge of the physical model in order to obtain a
mathematically rigorous context for prediction theory. Thus we are
able to state precisely what an optimal predictor should do. Unfortunately,
the predictors in current use are not optimal. 

The restricted setting where the physical model is assumed to be entirely
unknown is advantageous in making a connection to the fundamental
results of Wyner and Ziv \cite[1989]{WynerZiv1989} and Ornstein and
Weiss \cite[1992]{OrnsteinWeiss1992} in information theory. A quantity
that plays a central role in determining the predictability of a signal
is metric entropy. For examples where a physical model is useful for
prediction, see \cite{BrownEmery1998,ChernyshenkoBondarenko2008}.
As examples of chaotic signals whose entropy is not too high, we mention
signals obtained from Taylor-Couette flow and Rayleigh-Benard flow
\cite{FarmerSidorowich1987}. Turbulent signals in atmospheric boundary
layers, such as those recorded in  \cite{Narasimha2007}, have very
high entropy.

In the rest of this introduction, we summarize the contributions of
this paper and point out connections to other lines of research such
as data assimilation.

\begin{figure}
\begin{centering}
\includegraphics[scale=0.6]{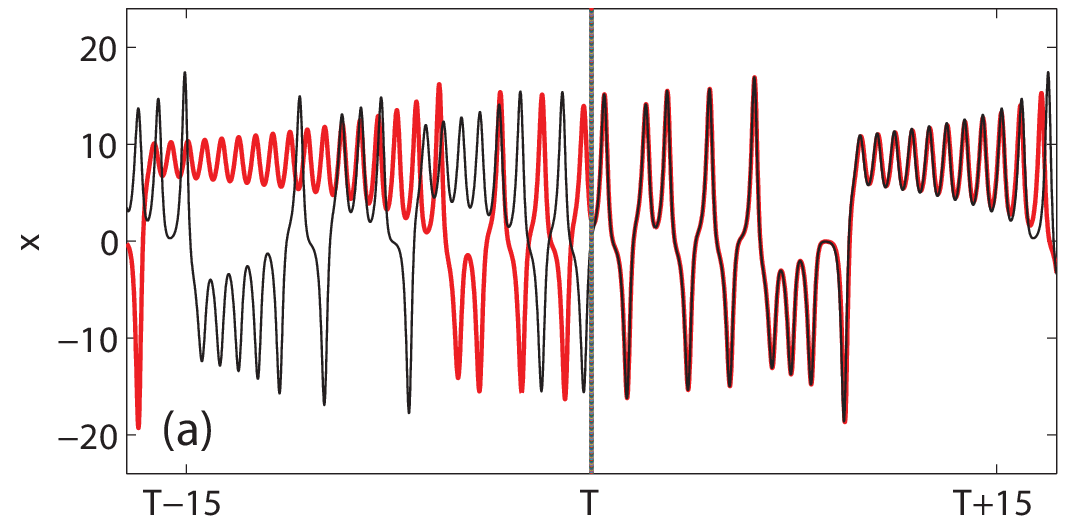}
\par\end{centering}

\begin{centering}
\includegraphics[scale=0.6]{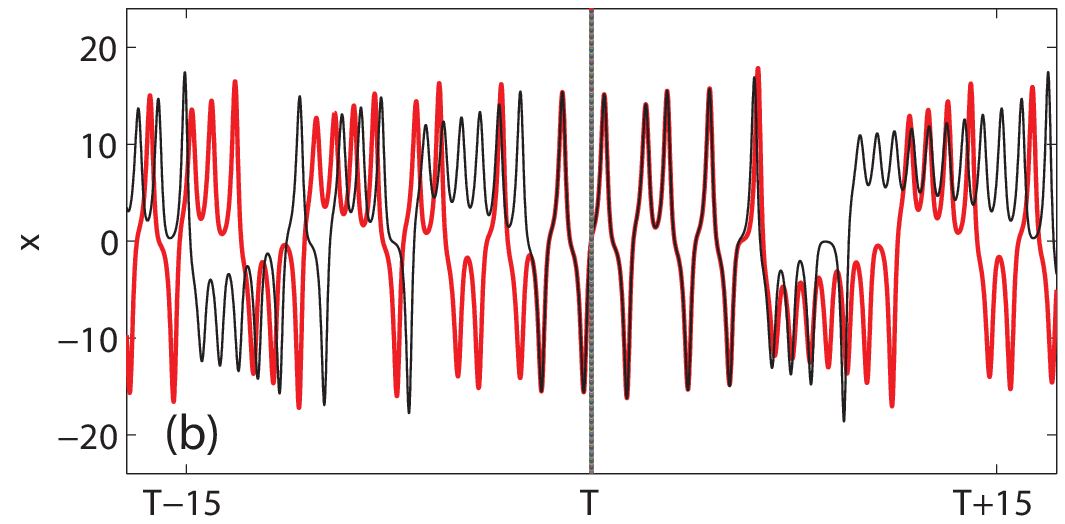}
\par\end{centering}

\caption{In both plots, the current time $T$ is $2^{20}$ symbols. The Lorenz
signal, which is shown as a thin black line, is the same in the two
plots. The thick red line is: (a) best fit from the past; (b) suboptimal
prediction using the embedding predictor.\label{fig:In-both-plots, Lorenz signal}}
\end{figure}

\textbf{Optimality and sub-optimality of prediction.} Perhaps the
central point of this paper is that matching ``the pattern of events''
is not the best way to predict chaotic signals in spite of its indubitable
intuitive appeal. This point is illustrated in Figure \ref{fig:In-both-plots, Lorenz signal}.
Before explaining that figure, we set down some notation that will
be used throughout this paper. If the signal is $x(t)$, the current
time is \emph{always }denoted by $T$. It is assumed that the signal
is recorded from $t=0$ and that the stretch of signal that is available
is $x(t)$ for $0\leq t\leq T$. The task is to use the available
history, which is $x(t)$ for $0\leq t\leq T$, to predict $x(T+t)$
for $0<t\leq t_{f}$ for as large a value of $t_{f}$ as possible. 

In the two plots of Figure \ref{fig:In-both-plots, Lorenz signal},
the thin black lines show a chaotic signal obtained from the Lorenz
system. The plots show only a part of the signal and $T$ is given
as $2^{20}$ symbols. Each symbol is equal to $t_{return}=0.7511$
units of time, where $t_{return}$ is the average time from one ``turning
point'' to another. A turning point is defined as a peak or a trough
of the graph of $x(t)$ with only peaks or troughs with $|x(t)|>6\sqrt{2}$
being counted. Since the two fixed points that are located in the
holes in the wings of the Lorenz attractor have coordinates $(\pm6\sqrt{2},$$\pm6\sqrt{2},27)$,
turning points defined in this way are in correspondence with intersections
of the signal with Poincaré sections of the Lorenz attractor \cite{Viswanath2004}.
In the plots, $T$ is given as $2^{20}$ symbols, which means $T=2^{20}\times0.7511$
and that the number of turning points of $x(t)$ in $[0,T]$ is approximately
$2^{20}$. In the plots, the thin black lines go beyond $T$ to show
how the signal develops so that we can visually assess the quality
of the predictions.

The thick red lines in the two plots are obtained differently. In
the top plot, we fix a tolerance $tol$ (the precise value of $tol$
is unimportant for the discussion here) and look for $t^{\ast}\in[0,T-t_{return}]$
such that the length of fit, which is 
\begin{equation}
\text{length of fit at }t^{\ast}=\text{largest}\: t_{f}\:\text{such that}\:|x(T+s)-x(t^{\ast}+s)|\leq tol\:\text{for}\: s\in[0,t_{f}],\label{eq:length-of-fit}
\end{equation}
is maximized. The maximum value of the length of fit is denoted by
$t_{best}$. Here we are looking into the future of the signal and
trying to find the moment $t^{\ast}$ in the past which agrees with
the signal's future for the maximum period $t_{f}$ (within a specified
tolerance). This method of determining $t^{\ast}$ and the maximum
length of fit $t_{best}$ will be called the best fit from the past.
Since it looks at $x(T+s)$ for $s>0$, the best fit from the past
is not a predictor. 

We see that the best fit from the past in Figure \ref{fig:In-both-plots, Lorenz signal}a
follows the signal for $t>T$ for more than $20$ symbols. It is not
difficult to see why no predictor can follow the signal for longer.
If we fit the signal starting at $x(t)$, $0\leq t\leq T-t_{return}$,
to the signal starting at $x(T)$ the fit will extend from $T$ to
$T+t_{f}$ for some $t_{f}$ and then start diverging. The rate of
divergence beyond $T+t_{f}$ will be exponential as the signal is
from a chaotic source. By definition of $t_{best}$, we have $t_{f}\leq t_{best}$.
Thus the past has no information about what happens to the signal
beyond $T+t_{best}$ and no amount of algorithmic legerdemain can
synthesize that information.

Currently available methods for prediction are based on recurrence
and embed the signal in phase space in one way or another \cite{Abarbanel1996,Boffetta2002,Casdagli1989,FarmerSidorowich1987,SugiharaMay1990}.
All the current predictors known to us use delay coordinates. Suppose
the real valued signal $x(t)$ is obtained as $x(t)=b^{T}X(t)$, where
$X(t)$ takes values in $\mathbb{R}^{d}$ and $b\in\mathbb{R}^{d}$
is constant. Suppose $X(t)$ satisfies the dynamical system $\dot{X}=f(X)$.
Delay coordinates are an attempt to reconstruct the dynamics of $X$
in $\mathbb{R}^{d}$ using the scalar signal $x(t)$. Even though
an individual signal value such as $x\left(t_{0}\right)$ may give
little idea of $X(t_{0})$, the pattern of events $x\left(t_{0}\right),x\left(t_{0}-\tau\right),\ldots x\left(t_{0}-(k-1)\tau\right)$
can be used to stand as a substitute for $X\left(t_{0}\right)$ and
to reconstruct dynamics in phase space for suitable values of the
delay $\tau$ and the embedding dimension $k$ \cite{EckmannRuelle1992,Takens1981}. 

It should be clear that the basic task of a predictor is to find $t^{\ast}$
which maximizes \prettyref{eq:length-of-fit} or another $t^{\ast}$
which nearly maximizes it without looking into the future. The embedding
predictors do not accomplish that task in an optimal way. For the
discussion here, a brief account of the basic embedding predictor
suffices. A more detailed discussion including extensions and modifications
of the basic predictor will be given later. The embedding predictor
works by finding $t^{\ast}\in[k\tau,T-t_{return}]$ such that 
\begin{equation}
\sum_{p=0}^{k-1}\left(x(T-p\tau)-x(t^{\ast}-p\tau)\right)^{2}\label{eq:embed-fit}
\end{equation}
is minimized. There is much literature about the choice of the delay
parameter $\tau$ and the embedding dimension $k$ (see \cite{Abarbanel1996}
for instance). We will assume that $\tau$ and $k$ are suitably chosen
(with $\tau k$ about a fifth of a symbol). The prediction of $x(T+s)$
is taken to be $x\left(t^{\ast}+s\right)$. 

How well does $t^{\ast}$ which minimizes \prettyref{eq:embed-fit}
work in terms of maximizing \prettyref{eq:length-of-fit}? Before
answering that question, let us ask ourselves why there should be
a connection at all between finding $t^{\ast}$ to minimize \prettyref{eq:embed-fit}
and finding it to maximize the length of fit defined by \prettyref{eq:length-of-fit}.
When we minimize \prettyref{eq:embed-fit}, we are looking for a $t^{\ast}$
such that if we walk back from $t=t^{\ast}$ the portion of the signal
we see looks much like what we see when we walk back from $t=T$.
In other words, \emph{the pattern of events }leading up to $t=t^{\ast}$
should look like the pattern of events leading up to $t=T$. The hope
is that if the events immediately preceding $t=t^{\ast}$ look like
the events immediately preceding $t=T$, the signal value $x(t^{\ast}+s)$
will be a good predictor of $x(T+s)$. 

Unfortunately, this intuitively reasonable idea is fundamentally deficient.
To see why, we go back to \prettyref{fig:In-both-plots, Lorenz signal}.
The thick red line of part (a) of that figure is obtained by shifting
$t^{\ast}$, which corresponds to the best fit from the past, to coincide
with $T$ to permit comparison between the two patterns. The thick
red line of part (b) is obtained by shifting $t^{\ast}$ found using
the embedding predictor to $T$. In part (a), we see that the sequence
of events leading up to $t=T$ and $t=t^{\ast}$ are not close at
all. Yet the two portions of the signal nearly converge at $T$ and
follow each other for more than twenty symbols. In part (b), on the
other hand, the sequence of events leading up $t=T$ and $t=t^{\ast}$
are actually quite close. If we walk backwards, the pattern of events
matches for three symbols. Yet the fit into the future is not half
as good as in part (a).

The situation shown in Figure \ref{fig:In-both-plots, Lorenz signal}
is typical. Because of the nature of chaotic signals, best fits tend
to converge at $t=T$ and diverge rapidly beyond $t=T+t_{best}$.
This introduces a fundamental asymmetry between the immediate past
and the immediate future. Good agreement in the immediate past does
not imply that the two portions of the signal will agree closely in
the future. 

Current predictors for predicting chaotic signals try to find a $t^{\ast}$
such that the pattern of events leading up to $t=t^{\ast}$ closely
resembles the pattern of events leading up to $t=T$. If the goal
is to predict the signal as far into the future as possible, that
is not the right idea. The right idea for an optimal predictor is
to evaluate if the pattern of events leading up to $t=t^{\ast}$ and
$t=T$ are such that the two patterns will come close to each other
in the future and to calculate for approximately how long they will
remain close. Such a calculation requires us to decompose the distance
between the two patterns into stable and unstable components. 

What are these stable and unstable components? Ideally, one would
like to define a notion of stable and unstable components that uses
signals and nothing more. Since no optimal general purpose predictor
of chaotic signals is currently known, such a notion cannot be made
precise. However, it is clear that such a notion has to correspond
in some way with stable and unstable manifolds or with stable and
unstable directions associated with local Lyapunov exponents of the
underlying dynamical system. 

In the limit of $T\rightarrow\infty$, the stable and unstable components
may be identified with the stable and unstable manifolds. However,
for finite $T$, especially considering the short intervals for which
prediction is possible, one has to use a notion of stable and unstable
components associated with local Lyapunov exponents. These fixed intervals
of time used for defining local Lyapunov exponents can be taken as
$\log_{2}T/H$. 

If we split the distance between the pattern of events leading up
to $t=T$ (black line in Figure \ref{fig:In-both-plots, Lorenz signal}a)
and the pattern of events leading up to $t=t^{\ast}$ for the best
fit (thick red line with $t^{\ast}$ shifted to $T$ in Figure \ref{fig:In-both-plots, Lorenz signal}a),
the distance between the two patterns has a noticeably substantial
stable component but a small unstable component. However, the stable
component decreases exponentially fast beyond $t=T$ which means the
two signals converge and stay close for an interval of time. The smallness
of the unstable component allows the fit between the two signals to
persist for the longest interval of time.

\textbf{Metric entropy.} Section 2 states a theorem of Wyner-Ziv \cite{WynerZiv1989}
and Ornstein-Weiss \cite{OrnsteinWeiss1992} and Sections 3 and 4
develop the implications of the entropy bound in that theorem to the
prediction of chaotic signals . Heuristically, the theorem says that
\[
\lim_{T\rightarrow\infty}\frac{t_{best}}{\log_{2}T}=\frac{1}{H}
\]
with probability 1. Here $H$ is entropy in a sense that will be described.
A predictor is optimal if it predicts the signal in the interval $\bigl[T,T+t_{f}\bigr)$
and 
\[
\liminf_{T\rightarrow\infty}\frac{t_{f}}{\log_{2}T}\geq\frac{1-\epsilon}{H}
\]
with probability 1 and for any $\epsilon>0$. In Section 5, we discuss
current predictors and point out why they are necessarily suboptimal.
In Sections 6 and 7, we develop a few ideas that take us closer to
a general purpose optimal predictor for chaotic signals. 

\textbf{Data assimilation and shadowing filters. }Shadowing filters
have been proposed as a method for state estimation and data assimilation.
Before discussing the connection of this paper to shadowing, we give
a brief discussion of data assimilation. This brief discussion has
two goals. It has been stated that ``the forecast skill of atmospheric
models depends not only on the accuracy of the initial conditions
and the realism of the model, but also on the instabilities of the
flow itself'' \cite[p. 227]{Kalnay2003}. This paper is focused exclusively
on the instabilities of the flow. Since weather and ocean models \cite{Bennett2002,Kalnay2003}
are major applications of prediction theory, it is perhaps not out
of place to call attention to measurement and modeling errors. Secondly,
the discussion provides some context for shadowing filters.

The following equation provides a framework for many data assimilation
techniques \cite{Kalnay2003}:
\[
X^{a}=X^{b}+W\left(Y^{o}-H\left(X^{b}\right)\right).
\]
In this equation, $X^{a}$ and $X^{b}$ are vectors which correspond
to a point in the state space of the physical model. Atmosphere model
variables typically include wind velocity components, temperature,
moisture, and surface pressure. The observation vector is denoted
by $Y^{0}$. Observed variables such as satellite radiances and radar
reflectivities do not occur in the physical model. The observation
operator $H$ maps the state vector of the physical model to observation
space. In weather prediction as well as climate modeling, the number
of degrees of freedom in the physical model is orders of magnitude
greater than the number of observations. Therefore it is impossible
to synthesize the current state of the model $X^{a}$ from observations
alone. The background field $X^{b}$ obtained from a short term forecast
is used as a starting point for inferring the current state of the
physical model.

The essence of data assimilation in this framework is the operator
$W$, which matches the observations against the background field
and generates a correction. Techniques such as optimal interpolation,
3DVar, and PSAS come under this framework. For a mathematical study
of such techniques, see \cite{apte2008data}. All practical methods
must account for the covariance of measurement error. In numerical
weather forecasting, this type of data assimilation is performed in
six hour cycles and information gradually propagates from regions
rich in observations to regions poor in observations \cite{Kalnay2003}. 

Another family of techniques explicitly allows observations to be
functions of time \cite{Kalnay2003}. One of these is the extended
Kalman filter. The extended Kalman filter updates the covariance matrix
of the estimated state vector from time to time using new observations.
Propagation and manipulation of the covariance matrix for realistic
physical models can be expensive. The ensemble Kalman filter is a
cheaper variant which introduces random errors into observations and
tracks several trajectories to estimate the covariance matrix. Yet
another technique is 4DVar. This technique finds the initial state
$X_{0}$ to minimize the quantity
\[
\left(X_{0}-X^{b}\right)^{T}B^{-1}\left(X-X_{0}\right)+\sum_{i=0}^{N}\left(H\left(X_{i}\right)-Y_{i}^{o}\right)^{T}R_{i}^{-1}\left(H\left(X_{i}\right)-Y_{i}^{o}\right).
\]
Here $R_{i}^ {}$is the covariance matrix of observations recorded
at time $t_{i}$. The state $X_{i}$ at time $t_{i}$ must be obtained
by integrating the physical model assuming the state at $t_{0}$ to
be $X_{0}$. The background field has a significant role in this technique
as well. The shadowing filters, to which we now turn, solve a minimization
problem that is formally similar to 4DVar.

The trajectory $\tilde{X}(t)$ is an $\epsilon$-orbit of the dynamical
system $dX/dt=f(X)$ if $||d\tilde{X}/dt-f(\tilde{X})||<\epsilon$
for all $t$. The shadowing lemma states that if the $\epsilon$-orbit
stays inside a suitable neighborhood of a hyperbolic invariant set
and for $\epsilon$ small enough, the $\epsilon$-orbit is $\delta$-shadowed
by a true orbit of the dynamical system \cite{Bowen1975,KatokHasselblatt1997}.
A similar result applies to hyperbolic invariant sets of maps. Hammel
et al. \cite{HammelYorkeGrebogi1988} have shown that the numerically
computed orbit of the Henon map $(u_{n+1},v_{n+1})=(1-Au_{n}^{2}+v_{n},-Ju_{n})$
with $A=1.4$, $J=-0.3$, and $(u_{0},v_{0})=(0,0)$ is $\delta$-shadowed
by a true orbit for up to $N=10^{7}$ iterations with $\delta=10^{-8}$,
even though the Henon map is not uniformly hyperbolic. A similar result
is given for the Ikeda map.

The essence of shadowing is that the error committed in each step
of an iteration may be decomposed along stable and unstable manifolds.
The error along the stable manifold can be canceled using a very small
perturbation at the final point of the trajectory. Similarly, the
error along the unstable manifold can be canceled using a very small
perturbation at the initial point of the trajectory. 

As already mentioned, shadowing filters have been proposed for noise
reduction and state estimation \cite{Hammel1990,FarmerSidorowich1991,judd2001indistinguishable,judd2004indistinguishable,stemler2009guide}.
For example, if $s_{t}$ are noisy observations of the state $x_{t}$
for $t=1,\ldots,T$, one may attempt to calculate the noise $\delta_{t}$
by minimizing $\sum_{t=1}^{T}||e_{t}||^{2}$, where $e_{t}=s_{t+1}-\delta_{t+1}-f(s_{t}-\delta_{t})$
\cite{judd2001indistinguishable}. Here the physical model $f$ is
assumed to be known. Other versions of the shadowing filter assume
only partial knowledge of $f$ \cite{judd2004indistinguishable}.

Judd and Smith \cite{judd2001indistinguishable} have considered various
gradient descent methods for denoising and stated a ``dictum'' based
on numerical experience. Their dictum is that the end point of the
estimated trajectory will lie close to the unstable manifold of the
end point of the true trajectory. In other words, much of the error
near the end point is along the unstable direction. Unfortunately,
this is the exact opposite of what an optimal predictor must do. This
situation results because the shadowing filter uses gradient descent
to match the entire segment of the trajectory as nearly as possible
and is therefore biased to fitting the past. The asymmetry between
fitting the past and predicting the future is not broken in favor
of the latter.

\textbf{Work of S.J. Lalley. }Lalley and others \cite{Lalley1999,lalley2006denoising}
have subjected the problem of denoising deterministic signals to an
incisive mathematical investigation. Lalley has found that an effective
de-noising algorithm must increase the width of the matching window
at a sub-logarithmic rate. Some of the considerations that led to
that finding could be relevant to optimal prediction. Lalley's work
provides a useful contrast to more applied work on de-noising. Judd
and Smith \cite{judd2004indistinguishable} refer to an earlier paper
of theirs and state, ``we showed, that contrary to what might be
expected, collecting more and more data will not provide a continually
improving estimate of the true state of a chaotic system.'' Here
it must be understood that Judd and Smith are referring to shadowing
filters based on gradient descent. In fact, Lalley \cite{Lalley1999}
has used very general assumptions on additive noise to prove that
his algorithm can recover the state of a chaotic system by collecting
more and more data.

\section{Theorem of Wyner-Ziv and Ornstein-Weiss}

In this section, we describe three results that apply to stationary
and ergodic sequences: the Poincaré recurrence theorem, a theorem
of Kac, and the entropy theorem of Wyner-Ziv and Ornstein-Weiss. Each
of these results is pertinent to source coding and, as we will show,
to the prediction of chaotic signals.

The notion of stationarity can be defined for a sequence of random
variables or for a dynamical system. Since our interest is in the
prediction of signals, we begin with the definition for a sequence
of random variables. A sequence of real valued random variables 
\[
X_{0},X_{1},X_{2},\ldots
\]
is stationary if 
\[
\mathbb{P}\left(\left(X_{n},X_{n+1},X_{n+2},\ldots\right)\in B\right)=\mathbb{P}\left(\left(X_{n+1},X_{n+2},X_{n+3}\ldots\right)\in B\right)
\]
for any Borel measurable subset $B$ of $\mathbb{R}^{\infty}$. The
definition captures the idea that the mechanism underlying the stochastic
process does not change with time. 

A stationary sequence is ergodic if every invariant event has probability
$0$ or $1$. Events phrased using means and correlations of the sequence
are examples of invariant events. 

For an alternative definition, let $T:\Omega\rightarrow\Omega$ be
a measurable transformation that preserves the probability measure
$\mu$ on $\Omega$. The set $A\subset\Omega$ is invariant if $T^{-1}A=A$.
The transformation $T$ is ergodic if $\mu(A)=0$ or $\mu(A)=1$ for
every invariant set $A$. The ergodicity condition precludes the dynamics
from getting stuck in a part of phase space. 

The Poincaré recurrence theorem does not assume ergodicity.
\begin{thm}[Poincaré recurrence \cite{KatokHasselblatt1997}]
 Assume $X_{0}$ to be $\mu$-distributed and define the stationary
sequence $X_{0},X_{1},\ldots$ with $X_{n}=T^{n}(X_{0})$ for $n=1,2,\ldots$
For a measurable subset $B$ of $\Omega$ with $\mu(B)>0$, $X_{0}\in B$
implies $X_{n}\in B$ infinitely often with probability 1. 
\end{thm}
Suppose a long stream of text is modeled as a stationary sequence
of characters and suppose that the set $B$ is chosen to prescribe
the first ten characters of the text. The theorem then asserts that
the sequence formed by the first ten characters will repeat again
and again. The origin of the sequence $X_{0}$ can be taken anywhere
in the text. 

If $\Omega$ is the phase space of a dynamical system, the theorem
asserts that the dynamical system will revisit the same region $B$
in phase space infinitely often. These revisitations are the basis
for predicting chaotic signals. 

The Poincaré recurrence is not quantitative. It does not tell us by
what factor a long stream of text can be compressed if the repetitions
are exploited or how well a chaotic signal can be predicted by tracking
the recurrences. The first step to a quantitative version of the Poincaré
recurrence theorem is a lovely theorem of Kac. This theorem assumes
the sequence to be ergodic.
\begin{thm}[Kac's theorem \cite{Kac1947}]
 Suppose that the sequence $X_{0},X_{1},\ldots$ is stationary and
ergodic. Let $B\subset\mathbb{R}$ with $\mathbb{P}(B)=\mathbb{P}\left(X_{0}\in B\right)>0$.
Let $n\geq1$ be the smallest integer such that $X_{n}\in B$. Then
$\mathbb{E}\left(n\bigl|X_{0}\in B\right)=1/\mathbb{P}(B)$.\label{thm:Kac}
\end{thm}
Kac's theorem says that the expected time to return to the set $B$
is exactly equal to the inverse of the probability of $B$. One would
expect the recurrence time to sets of smaller probability to be greater.
While the elegance of Kac's theorem may lead one to suspect that the
theorem should be obvious or easy to demonstrate, a perusal of Kac's
ingenious proof will dispel such a misperception.

The entropy theorem stated below characterizes recurrences more sharply
than Kac's theorem. It applies to sequences which are stationary,
ergodic, and take values in a finite alphabet. The restriction to
finite alphabets does not cause such a great loss of generality because
information is fundamentally discrete in nature. Chaotic signals are
real valued and often continuous in time. Yet we may obtain a notion
of optimality of prediction of chaotic signals using the entropy theorem,
as we will show in the following sections. 

Since $X_{n}$ is assumed to take values in a finite alphabet $\mathcal{A}$
for $n\geq0$, we refer to each value as a symbol. The entropy $H$
is defined as follows. Suppose we consider the following block of
symbols of length $m$: $X_{0},\ldots X_{m-1}$. This block can take
$|\mathcal{A}|^{m}$ different values. Suppose the probabilities of
the different possibilities are $p_{1},p_{2},\ldots p_{M}$, where
$M=|\mathcal{A}|^{m}$. Then 
\[
H=\lim_{m\rightarrow\infty}\frac{1}{M}\sum_{i=1}^{M}-p_{i}\log_{2}p_{i}.
\]
 We will follow the information theory convention and use logarithms
to base $2$.

The definition of entropy comes up in a natural way when we try to
count states. Suppose we look at all $|\mathcal{A}|^{m}$ possible
values of the sequence $X_{0},\ldots X_{m-1}$. Some possible sequences
are more probable and some are less probable. How many possible sequences
have a probability that is approximately that of the average? The
answer is $2^{mH}$. The entropy theorem of Shannon and others asserts
that a sufficiently long segment of $X_{0},X_{1},\ldots$ looks like
an average sequence with probability $1$. Therefore to transmit $m$
symbols from such a stationary and ergodic source, we may be able
to get by using $mH$ bits but no less. An optimal compression of
the source will use $mH$ bits to encode $m$ symbols asymptotically.
\begin{thm}[Ornstein and Weiss \cite{OrnsteinWeiss1992}]
 Let $X_{0},X_{1},\ldots$ be a stationary and ergodic sequence,
in which each $X_{n}$ takes values in a finite alphabet $\mathcal{A}$.
Let $t_{best}$ be the greatest integer such that $X_{T+1},\ldots,X_{T+t_{best}}$
occurs as a contiguous subsequence of $X_{0},\ldots,X_{T}$. Then
\[
\lim_{T\rightarrow\infty}\frac{t_{best}}{\log_{2}T}=\frac{1}{H}
\]
with probability 1. Here $H$ is the entropy of the stationary, ergodic
process $X_{0},X_{1},\ldots$\label{thm:WynerZiv-OrnsteinWeiss}
\end{thm}
Theorem \ref{thm:Kac} tracks the re-occurrence of an event associated
with $X_{0}$ for some $X_{n}$ with $n>0$. Theorem \ref{thm:WynerZiv-OrnsteinWeiss}
checks if an event that follows the current symbol $X_{T}$ repeats
a past event. We will refer to either scenario as a recurrence. 

Theorem \ref{thm:WynerZiv-OrnsteinWeiss} is a remarkable sharpening
of the Poincaré recurrence theorem. If we regard $T$ as current time
and that observations begin at $0$, as we do throughout this paper,
it gives a perfect characterization of the extent to which the pattern
that will follow $T$ will repeat some pattern of events we have seen
in the past. The $1/H$ bound was first stated by Wyner and Ziv \cite{WynerZiv1989},
who were able to prove the convergence of $t_{best}/\log_{2}T$ to
$1/H$ in probability. Almost sure convergence of the type asserted
by Theorem \ref{thm:WynerZiv-OrnsteinWeiss} was proved by Ornstein
and Weiss \cite{OrnsteinWeiss1992}. 

The distinction between convergence in probability and almost sure
convergence is pertinent to the prediction of chaotic signals. If
predictions of weather or of hurricane tracks or of cardiac signals
are to be really useful, the prediction method should apply to almost
every signal and not only to a fraction of the signals that occur
in practice. The distinction between almost sure predictions of individual
signals and statistical predictability has not been made in extant
work on the subject. Existing predictors of chaotic signals have been
validated generally with statistical notions of accuracy such as mean
square error and correlation plots \cite{FarmerSidorowich1987,SugiharaMay1990}.
Our discussion of predictability of chaotic signals will be framed
in terms of almost sure predictability. 

Entropy comes up in statistical mechanics while counting the number
of states of a system. The interpretation of entropy in terms of information
originated with Shannon's source coding theorem. However, the coding
scheme implicit in Shannon's theorem, which is to use long block codes,
is useless in practice. The widely used source coding scheme of Lempel
and Ziv relies on an entirely different interpretation of entropy,
which is the interpretation given by Theorem \ref{thm:WynerZiv-OrnsteinWeiss}. 

Theorem \ref{thm:WynerZiv-OrnsteinWeiss} interprets entropy in terms
of the maximum segment following $X_{T}$ that occurs as a contiguous
subsequence of the segment preceding it. The entire segment following
$X_{T}$ can be encoded using a pointer to some place in the past
and the length of the recurrence. Various source coding schemes based
on that idea have been derived by Lempel, Ziv and others and have
been widely used for decades. The distinction between almost sure
convergence and convergence in probability is important for the practical
success of these coding schemes.

In normal use, entropy theorem refers to the entropy theorem of Shannon.
In this paper, entropy theorem and entropy bound will refer to Theorem
\ref{thm:WynerZiv-OrnsteinWeiss}. This convention saves us the trouble
of using four names every time we need to refer to the theorem and
the bound contained in it.

If we look at the entropy theorem in the light of prediction, it appears
as if $\log_{2}T/H$ symbols can be predicted using a history of length
$T$. The fallacy behind that surmise becomes evident if we consider
an i.i.d. sequence made up of $\pm1$, where each sign is equally
probable. The entropy of such a sequence is $1$. As the entropy theorem
asserts, we may expect $\log_{2}T$ symbols that follow a history
of length $T$ to form a segment that repeats a segment that has already
occurred. That type of repetition is useless for prediction. Given
a knowledge of the history of the signal up to $X_{T}$, all that
we know about $X_{T+1}$ is that it is equally likely to be $+1$
or $-1$. Knowledge of history is useless in the prediction of i.i.d.
sequences.

Thus we need to precisely delineate the nature of chaotic signals
which makes the entropy theorem relevant to their prediction. In Section
3, we describe the notion of entropy for chaotic signals, and in Section
4, we  explain why the entropy theorem defines the limit of predictability
of chaotic signals. At the end of Section 6, we describe what form
on optimal predictor should take. While currently available predictors
do not take that form, in the rest of the paper, we describe a few
ideas that suggest that optimal predictors can in fact be derived.

\section{Applicability of the entropy theorem to chaotic systems}

Stationary and ergodic sequences can be generated in many ways. An
i.i.d. sequence $X_{0}=\pm1,\: X_{1}=\pm1,\ldots$ with $p(+1)=p(-1)=1/2$
is stationary and ergodic. Suppose we form another sequence $Y_{n}$
with $Y_{n}=1$ or $Y_{n}=-1$ according as $+1$ or $-1$ is the
majority among the seven entries $X_{n},\ldots,X_{n+6}$. Then the
$Y_{n}$ sequence is also stationary and ergodic. Regardless of the
length of history neither the $X_{n}$ sequence nor the $Y_{n}$ sequence
is predictable in the manner we consider. For notions of prediction
pertinent to such signals, see \cite{Furstenberg1960}.

Suppose $X_{n+1}=f\left(X_{n}\right)$ is a dynamical system. The
phase space of the dynamical system can be any Riemannian manifold
but for convenience we will assume it to be a subset of $\mathbb{R}^{d}$.
Let $\mu$ be a probability measure that is invariant with respect
to the dynamical system (in other words $\mu(A)=\mu(f^{-1}(A))$ for
Borel sets A). If $X_{0}$ has $\mu$ as its distribution and $X_{n+1}=f\left(X_{n}\right)$
for $n=0,1,\ldots$, the sequence $X_{0},X_{1},\ldots$ is stationary.
If $\mu$ is indecomposable (an assumption we will always make), the
sequence is ergodic as well.

It is evident that a stationary and ergodic sequence $X_{0},X_{1},\ldots$
generated in this manner is quite different from an i.i.d. sequence
of the type $\pm1,\pm1,\ldots$ While the i.i.d. sequence generates
a random number for every new entry, in a stationary and ergodic sequence
derived from a dynamical system, every new entry is generated deterministically. 

We do not assume the entire state vector $X_{n}$ to be observable.
The observed sequence is $x_{0},x_{1},\ldots$ where $x_{n}$ is a
real-valued function of $X_{n}$. For example, $x_{n}$ can be some
component of $X_{n}$. This framework should be sufficiently general
to allow for seismic signals, ECG signals and so on. Nearly all the
theoretical discussion will be restricted to maps to avoid some of
the technicalities that arise for flows. For both maps and flows,
the dynamical system that generates the signal is assumed to be unknown.

One of the examples we consider is a signal obtained from the Lorenz
flow:
\begin{eqnarray*}
\frac{dx}{dt} & = & 10(y-x)\\
\frac{dy}{dt} & = & 28x-y-xz\\
\frac{dz}{dt} & = & -8z/3+xy.
\end{eqnarray*}
The Lorenz system has fixed points at $(0,0,0)$ and $(\pm6\sqrt{2},\pm6\sqrt{2},27)$.
The two nonzero fixed points sit in the middle of holes in the two
wings of the butterfly-shaped attractor. The signal is generated by
accurately integrating a random point $\left(x',y',z'\right)$ for
some time to generate the initial point $(x(0),y(0),z(0))$. The initial
point generated in this way may be assumed to be $\mu$ distributed,
with $\mu$ assumed to be the physical measure of the Lorenz attractor.
The signal $x(t)$ is generated for $t\geq0$ by integrating this
initial point. For the purpose of prediction, it is assumed that the
model which generates the signal is unknown.

To apply the entropy theorem to the Lorenz signal $x(t)$, we need
to specify the entropy of the Lorenz signal. We recall a few of the
theoretical definitions related to the entropy of a dynamical system.
For complete details, see \cite{KatokHasselblatt1997} or \cite{Young2002}.
Let $f:\mathbb{R}^{d}\rightarrow\mathbb{R}^{d}$ be a smooth dynamical
system and let $\mathcal{A}$ be an invariant set. Let $\mu$ be a
probability measure on $\mathcal{A}$ that is invariant with respect
to $f$. Assume that $f$ is an ergodic transformation of $\mathcal{A}$
with respect to the measure $\mu$. In this setting, the definition
of metric or Kolmogorov-Sinai entropy is quite simple. Let $\mathcal{P}$
be a finite partition of the set $\mathcal{A}$. We can generate a
finite-valued stationary ergodic process as follows. Pick $X_{0}$
according to $\mu$ and take $X_{n+1}=f\left(X_{n}\right)$ for $n=0,1,\ldots$
Let $Y_{n}$ be the partition in $\mathcal{P}$ that $X_{n}$ belongs
to. Then the finite valued process $Y_{n}$ is stationary and ergodic,
and as such has a Shannon entropy which we denote by $h_{\mu}(f,\mathcal{P})$.
In general, $h_{\mu}\left(f,\mathcal{P}\right)$ can depend upon the
partition $\mathcal{P}.$ The metric entropy $h_{\mu}(f)$ is defined
as the maximum over all finite partitions $\mathcal{P}$.

At first sight, it might seem as if the dependence of metric entropy
on $\mathcal{P}$ could be a problem. However, this dependence is
not as severe as one might think. For example, one may modify $\mathcal{P}$
to the finer partition $\mathcal{P}\vee\mathcal{P}$, where the finer
partition keeps track of the partitions in $\mathcal{P}$ that $x$
and its iterate $f(x)$ belong to. Even though $\mathcal{P}\vee\mathcal{P}$
is a finer partition, $h_{\mu}\left(f,\mathcal{P}\vee\mathcal{P}\right)=h_{\mu}(f,\mathcal{P})$
because it is readily evident that combining the $n$-th and the $(n+1)$-st
symbols into a single symbol in the $n$-th position will neither
increase nor decrease the information per symbol. In fact, \texttt{$h_{\mu}(f)=h_{\mu}\left(f,\mathcal{P}\right)$}
if the partition $\mathcal{P}$ is \emph{generating}. Generating partitions
are defined using conditional entropy \cite{KatokHasselblatt1997}.
If the partitions in $\mathcal{P}\vee\ldots\vee\mathcal{P}$ become
fine enough to closely approximate any given partition $\mathcal{Q}$
of $\mathcal{A}$, the partition $\mathcal{P}$ is generating. 

Later the theoretical discussion will focus on hyberbolic attractors
$\mathcal{A}$. For such invariant sets, Markov partitions are generating.
But now we will explain how the concept of metric entropy allows us
to apply the entropy theorem to Lorenz signals.

\begin{table}
\begin{centering}
{\footnotesize }%
\begin{tabular}{|r|r|l||l|l|l|}
\hline 
{\footnotesize $\log_{2}T$} & {\footnotesize $t_{best}$} & {\footnotesize Matching sequence} & {\footnotesize $\log_{2}T$} & {\footnotesize $t_{best}$} & {\footnotesize Matching sequence}\tabularnewline
\hline 
\hline 
{\footnotesize 2} & {\footnotesize 2} & {\tiny BA} & {\footnotesize 12} & {\footnotesize 13} & {\tiny AAAABAAABBAAA}\tabularnewline
\hline 
{\footnotesize 3} & {\footnotesize 3} & {\tiny BBB} & {\footnotesize 13} & {\footnotesize 12} & {\tiny BBBBABAAABAA}\tabularnewline
\hline 
{\footnotesize 4} & {\footnotesize 7} & {\tiny ABABBBB} & {\footnotesize 14} & {\footnotesize 16} & {\tiny ABBBABBAAAAAAABA}\tabularnewline
\hline 
{\footnotesize 5} & {\footnotesize 7} & {\tiny BBAABAA} & {\footnotesize 15} & {\footnotesize 17} & {\tiny BBBABBBABABBAABAB}\tabularnewline
\hline 
{\footnotesize 6} & {\footnotesize 9} & {\tiny BBBBBBBAA} & {\footnotesize 16} & {\footnotesize 16} & {\tiny BAABABBBAAAAAABB}\tabularnewline
\hline 
{\footnotesize 7} & {\footnotesize 10} & {\tiny AAAAAAAAAA} & {\footnotesize 17} & {\footnotesize 20} & {\tiny AAABBBAAABABAAABAAAA}\tabularnewline
\hline 
{\footnotesize 8} & {\footnotesize 14} & {\tiny AABAABAAAAAAAA} & {\footnotesize 18} & {\footnotesize 24} & {\tiny BABBABBBABBBBABBABBABABB}\tabularnewline
\hline 
{\footnotesize 9} & {\footnotesize 9} & {\tiny BBBAABBAA} & {\footnotesize 19} & {\footnotesize 20} & {\tiny BBBBABABBBAABAAAABAA}\tabularnewline
\hline 
{\footnotesize 10} & {\footnotesize 10} & {\tiny BABABBBBAB} & {\footnotesize 20} & {\footnotesize 16} & {\tiny ABBBAABABAAAABA}\tabularnewline
\hline 
{\footnotesize 11} & {\footnotesize 9} & {\tiny ABBBAAAAA} & {\footnotesize 21} & {\footnotesize 30} & {\tiny BABBBBAABABBAAAABBBBAAAAAABABB}\tabularnewline
\hline 
\end{tabular}
\par\end{centering}{\footnotesize \par}

\caption{Recurrences of a Lorenz signal. The subsequence extending from position
$T+1$ to position $T+t_{best}$ is matched by a subsequence extending
from from position $t^{\ast}$ to $t^{\ast}+t_{best}$ , where $t^{\ast}$
is a position in the past with $t^{\ast}+t_{best}\leq T$. \label{tab:Lorenz-Wyner-Ziv}}
\end{table}

Table \ref{tab:Lorenz-Wyner-Ziv} shows a calculation of $t_{best}$,
in accord with its definition in the entropy theorem (Theorem \ref{thm:WynerZiv-OrnsteinWeiss}),
using a Lorenz signal. The symbols $A$ and $B$ have the following
meaning. Every intersection of the Lorenz signal with the ``quarter''
plane $x<-6\sqrt{2},\: y<-6\sqrt{2},\, z=27$ is recorded as the symbol
$A$ and every intersection with $x>6\sqrt{2},\, y>6\sqrt{2},\, z=27$
is recorded as the symbol $B$. In this manner the Lorenz signal is
turned into a stationary and ergodic sequence of $A$s and $B$s.
For evidence that the partition into $A$ and $B$ is generating,
see \cite{viswanath2003symbolic,Viswanath2004}.

A convenient way to estimate the entropy of the sequence of $A$s
and $B$s is using Lyapunov exponents. Lyapunov exponents are the
exponential rates with which infinitesimal perturbations to a point
on $\mathcal{A}$ grow or decay. For a definition, see \cite{KatokHasselblatt1997}.
The standard definition uses natural logarithms and not logarithms
to base $2$ as in the case of entropy. If the system is of dimension
$d$, there are exactly $d$ Lyapunov exponents counting multiplicities.
With probability $1$ with respect to the measure $\mu$, these are
the only possible rates of growth or decay. 

If the Lyapunov exponents are $\lambda_{1},\ldots,\lambda_{d}$, the
metric entropy satisfies 
\begin{equation}
h_{\mu}\leq\sum_{\lambda_{i}>0}\lambda_{i}.\label{eq:ruelle-inequality}
\end{equation}
This is Ruelle's inequality \cite{Young2002} (the same logarithm
must be used in defining $h_{\mu}$ and the Lyapunov exponents $\lambda_{i}$).
In some cases, equality holds in \prettyref{eq:ruelle-inequality}. 

For the Lorenz system, the continuous time Lyapunov exponent is approximately
$0.905$ (using natural logarithms). The average time from an intersection
with one of the quarter-planes $A$ or $B$ to another is $t_{return}=0.7511$.
By Ruelle's inequality \prettyref{eq:ruelle-inequality}, the entropy
of the sequences of $A$s and $B$s is bounded above by $0.905\times0.7511/\log2=0.98$.
The entropy appears to be close to $0.98$ \cite{viswanath2003symbolic,Viswanath2004}.
Table \ref{tab:Lorenz-Wyner-Ziv} appears to be in agreement with
this estimate of the entropy.

\begin{table}
\begin{centering}
{\footnotesize }%
\begin{tabular}{|r|r|l||l|l|l|}
\hline 
{\footnotesize $\log_{2}T$} & {\footnotesize $t_{best}$} & {\footnotesize Matching sequence} & {\footnotesize $\log_{2}T$} & {\footnotesize $t_{best}$} & {\footnotesize Matching sequence}\tabularnewline
\hline 
\hline 
{\footnotesize 2} & {\footnotesize 6} & {\tiny AABAAA} & {\footnotesize 12} & {\footnotesize 12} & {\tiny AAABAABAABBB}\tabularnewline
\hline 
{\footnotesize 3} & {\footnotesize 4} & {\tiny AAAB} & {\footnotesize 13} & {\footnotesize 12} & {\tiny BABBAAAAAABA}\tabularnewline
\hline 
{\footnotesize 4} & {\footnotesize 8} & {\tiny ABAAAABB} & {\footnotesize 14} & {\footnotesize 13} & {\tiny BBAABBBABABBB}\tabularnewline
\hline 
{\footnotesize 5} & {\footnotesize 7} & {\tiny BBABBAB} & {\footnotesize 15} & {\footnotesize 14} & {\tiny BBAAAAAAAAAAAB}\tabularnewline
\hline 
{\footnotesize 6} & {\footnotesize 6} & {\tiny BABAAA} & {\footnotesize 16} & {\footnotesize 16} & {\tiny BAAABBBABBBABABB}\tabularnewline
\hline 
{\footnotesize 7} & {\footnotesize 11} & {\tiny AABAAABBBBB} & {\footnotesize 17} & {\footnotesize 20} & {\tiny AAAABABABAABAABABBAB}\tabularnewline
\hline 
{\footnotesize 8} & {\footnotesize 5} & {\tiny BABBB} & {\footnotesize 18} & {\footnotesize 17} & {\tiny BAABAAAABAABAABBB}\tabularnewline
\hline 
{\footnotesize 9} & {\footnotesize 11} & {\tiny BBBBBBBABBA} & {\footnotesize 19} & {\footnotesize 22} & {\tiny BABBAAAAAABBBBAAABAAAB}\tabularnewline
\hline 
{\footnotesize 10} & {\footnotesize 12} & {\tiny BAAAABABAABA} & {\footnotesize 20} & {\footnotesize 19} & {\tiny BABAAAAABABABBABBAB}\tabularnewline
\hline 
{\footnotesize 11} & {\footnotesize 10} & {\tiny BABBABAABB} & {\footnotesize 21} & {\footnotesize 20} & {\tiny ABABBBBBBBBBABABBAAA}\tabularnewline
\hline 
\end{tabular}
\par\end{centering}{\footnotesize \par}

\caption{Recurrences of flips of a fair coin calculated in the same manner
as in the previous table.\label{tab:fair-coin-Wyner-Ziv}}
\end{table}

Table \ref{tab:fair-coin-Wyner-Ziv} tabulates $t_{best}$ (defined
as in Theorem \ref{thm:WynerZiv-OrnsteinWeiss}) for tosses of a fair
coin (with $A$ for heads and $B$ for tails). The entropy of the
coin toss process is $1$ and very close to the entropy of the Lorenz
signal. Yet Table \ref{tab:fair-coin-Wyner-Ziv} looks quite different
from Table \ref{tab:Lorenz-Wyner-Ziv}. The fluctuations of $t_{best}$
are more pronounced for the Lorenz signal. For the special case of
i.i.d. sequences such as coin tosses, Theorem \ref{thm:WynerZiv-OrnsteinWeiss}
was proved by Erdos and Renyi. 

\begin{figure}
\begin{centering}
\includegraphics[scale=0.6]{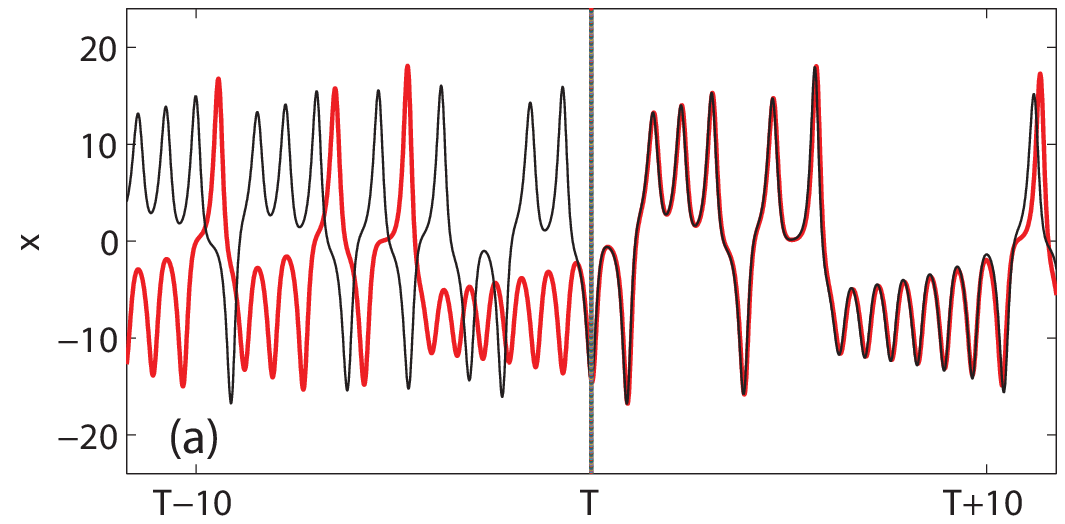}
\par\end{centering}

\centering{}\includegraphics[scale=0.6]{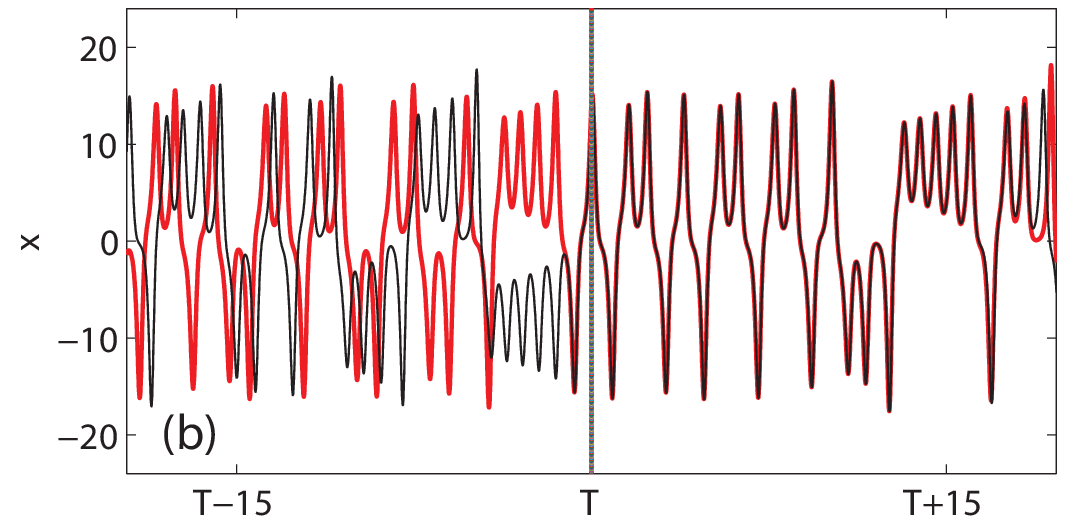}\caption{Best fits from the past (in thick red) to a Lorenz signal (in thin
black). (a) $T=2^{14}$ symbols. (b) $T=2^{21}$ symbols.\label{fig:Lorenz-best-fit} }
\end{figure}
The intersection with the quarter-planes $A$ and $B$ are recorded
using the symbols $A$ and $B$. For continuous time Lorenz signals
$x(t)$, one may use the the average time between symbols $t_{return}=0.7511$
as the unit. Following that usage, the value of the current time $T$
for the two plots in Figure \ref{fig:Lorenz-best-fit} are reported
as $2^{14}$ and $2^{21}$ symbols.

{\footnotesize }
\begin{table}
\begin{centering}
{\footnotesize }%
\begin{tabular}{|>{\centering}p{0.75in}|>{\centering}p{0.75in}|>{\centering}p{0.7in}||>{\centering}p{0.75in}|>{\centering}p{0.75in}|>{\centering}p{0.7in}|}
\hline 
{\footnotesize $\log_{2}T$}{\footnotesize \par}

{\footnotesize (in symbols)} & {\footnotesize $t_{best}$}{\footnotesize \par}

{\footnotesize (in symbols)} & {\footnotesize $t_{best}$ }{\footnotesize \par}

{\footnotesize (as a real)} & {\footnotesize $\log_{2}T$ }{\footnotesize \par}

{\footnotesize (in symbols)} & {\footnotesize $t_{best}$}{\footnotesize \par}

{\footnotesize (in symbols)} & {\footnotesize $t_{best}$ }{\footnotesize \par}

{\footnotesize (as a real)}\tabularnewline
\hline 
\hline 
{\footnotesize 2} & {\footnotesize 1} & {\footnotesize 0.58} & {\footnotesize 12} & {\footnotesize 9} & {\footnotesize 6.54}\tabularnewline
\hline 
{\footnotesize 3} & {\footnotesize 3} & {\footnotesize 2.20} & {\footnotesize 13} & {\footnotesize 13} & {\footnotesize 9.84}\tabularnewline
\hline 
{\footnotesize 4} & {\footnotesize 8} & {\footnotesize 6.04} & {\footnotesize 14} & {\footnotesize 15} & {\footnotesize 11.02}\tabularnewline
\hline 
{\footnotesize 5} & {\footnotesize 6} & {\footnotesize 4.81} & {\footnotesize 15} & {\footnotesize 13} & {\footnotesize 9.60}\tabularnewline
\hline 
{\footnotesize 6} & {\footnotesize 6} & {\footnotesize 4.78} & {\footnotesize 16} & {\footnotesize 17} & {\footnotesize 12.41}\tabularnewline
\hline 
{\footnotesize 7} & {\footnotesize 8} & {\footnotesize 5.97} & {\footnotesize 17} & {\footnotesize 18} & {\footnotesize 13.59}\tabularnewline
\hline 
{\footnotesize 8} & {\footnotesize 4} & {\footnotesize 3.04} & {\footnotesize 18} & {\footnotesize 14} & {\footnotesize 10.54}\tabularnewline
\hline 
{\footnotesize 9} & {\footnotesize 8} & {\footnotesize 5.85} & {\footnotesize 19} & {\footnotesize 18} & {\footnotesize 13.22}\tabularnewline
\hline 
{\footnotesize 10} & {\footnotesize 9} & {\footnotesize 6.69} & {\footnotesize 20} & {\footnotesize 22} & {\footnotesize 16.47}\tabularnewline
\hline 
{\footnotesize 11} & {\footnotesize 8} & {\footnotesize 5.92} & {\footnotesize 21} & {\footnotesize 25} & {\footnotesize 18.91}\tabularnewline
\hline 
\end{tabular}
\par\end{centering}{\footnotesize \par}

{\footnotesize \caption{Best fits to a Lorenz signal, where $t_{best}$ in symbols equals
$t_{best}$ as a real number divided by $t_{return}=0.7511$.\label{tab:Lorenz-best-fit}}
}
\end{table}
{\footnotesize \par}

When we think of the Lorenz signal as a sequence made up of the symbols
$A$ and $B$, it is natural to define $t_{best}$ as in the entropy
theorem (Theorem \ref{thm:WynerZiv-OrnsteinWeiss}). However, for
continuous time signals the definition of $t_{best}$ which follows
\prettyref{eq:length-of-fit} is more natural. We take 
\begin{equation}
tol=5\label{eq:lorenz-tolerance}
\end{equation}
to be the tolerance for Lorenz signals throughout this paper. Table
\ref{tab:Lorenz-best-fit} reports $t_{best}$ with $tol=5$. The
$t_{best}$ numbers with $tol=5$ are somewhat smaller than the $t_{best}$
numbers in Table \ref{tab:Lorenz-Wyner-Ziv}. This is because $tol=5$
is a stiffer requirement than simply requiring the symbol sequences
to match. When other methods are compared to the best fits in Table
\ref{tab:Lorenz-best-fit} later, the length of match is reported
in symbols but not as a real number.

\begin{figure}
\begin{centering}
\includegraphics[scale=0.4]{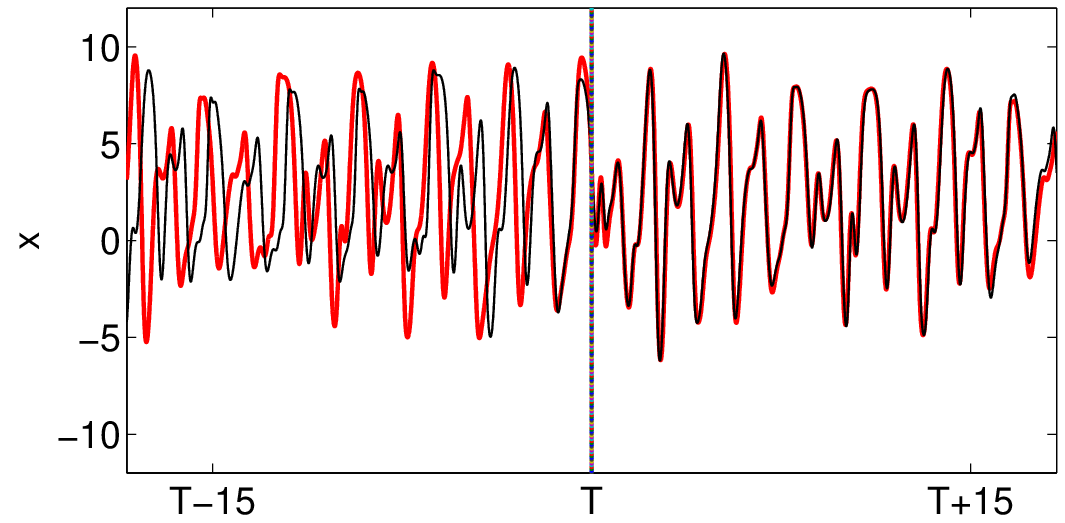}\hspace{0.5cm}%
\begin{tabular}[b]{|c|c|}
\hline 
$T$ & $t_{f}$\tabularnewline
\hline 
$10^{2}$ & $3.6$\tabularnewline
\hline 
$10^{3}$ & $7.4$\tabularnewline
\hline 
$10^{4}$ & $14.2$\tabularnewline
\hline 
$10^{5}$ & $18.4$\tabularnewline
\hline 
\end{tabular}
\par\end{centering}

\caption{Best fit from the past for a chaotic signal obtained from \eqref{eq:lorenz96}
using $f=8.17$ and a table showing the logarithmic dependence of
$t_{f}$ on $T$.\label{fig:lorenz96}}
\end{figure}
For another example of the applicability of the entropy theorem, we
turn to the following equations:
\begin{equation}
\frac{dx_{j}}{dt}=x_{j-1}\left(x_{j+1}-x_{j-2}\right)-x_{j}+f\label{eq:lorenz96}
\end{equation}
with $j=0,1,2,3,4$ and with the arithmetic in the subscripts being
modulo $5.$ When $f=8.17$ this system is chaotic \cite{abarbanel2009dynamical}.
As shown in Figure \ref{fig:lorenz96}, the entropy theorem applies
to this chaotic system (the signal is from $x_{0}$). As expected,
the best fit into the future diverges from the signal as we walk back
in time. This example was introduced in \cite{abarbanel2009dynamical}
to show that a single signal cannot be used to synchronize a chaotic
physical model. In this case, the system has two positive conditional
Lyapunov exponents and two signals are needed to synchronize the physical
model. The same point has come up in the theory of the Navier-Stokes
equations. For example, less than 5\% of the modes suffice to synchronize
turbulent channel flow but less than 1\% will not do \cite{ChernyshenkoBondarenko2008}.
The master modes or the determining modes must be sufficiently numerous
to capture the entire system, a point we alluded to at the beginning
of the introduction.

\section{Recurrence of chaotic signals and limits of predictability}

Suppose we are trying to predict a signal $x_{0},\ldots,x_{T}.$ The
entropy theorem says that $t_{best}\approx\log_{2}T/H$ for large
$T$. Thus it appears the past of the signal does not have sufficient
information to predict $x_{T+t}$ for $t>\log_{2}T/H$. We expect
that no algorithm can predict $x_{T+t}$ for $t>(1+\epsilon)\log_{2}T/H$
for $\epsilon>0$. In this section, we formalize this claim to some
extent to bring out in outline what form the proof of such a claim
might take.

As in Section 3, the sequence $x_{0},\ldots,x_{T}$ is assumed to
be generated from the state vectors of a dynamical system $X_{n+1}=f(X_{n})$.
Since our aim is to upper bound the extent of predictability of the
sequence, we may, without loss of generality, assume the entire state
vector to be observable. We assume that the map $f$ possesses a hyperbolic
attractor $\mathcal{A}$. We assume that $f$ is transitive on $\mathcal{A}$.
Within a hyperbolic attractor, periodic points are dense and therefore
a hyperbolic attractor satisfies the Axiom-A conditions. 

We define a predictor as a measurable function and write it as 
\[
P\left(X_{0},\ldots,X_{T}\right)=\left(\tilde{X}_{T+1},\tilde{X}_{T+2},\ldots\right).
\]
The measurable function $P$ captures our notion of an algorithm which
will take the $T$ successive state vectors $X_{0},\ldots,X_{T}$
and will generate approximations $\tilde{X}_{T+s}$ to $X_{T+s}$
for $s=1,2,\ldots$ The algorithm is not required to output an approximation
for every $s>0$. We will assume that it outputs approximations for
$s=1,2,\ldots,t_{f}$.

At this point, we have to decide when a prediction is termed as valid.
A prediction $\tilde{X}_{T+s}$ is deemed to be valid if $\bigl|X_{T+s}-\tilde{X}_{T+s}\bigr|\leq tol$
for some tolerance $tol$. We require the prediction algorithm to
output $\tilde{X}_{t+s}$ as a valid prediction for $s=1,\ldots,t_{f}$.
In other words, each prediction output by the prediction algorithm
$P$ must be valid. Alternatively, we can allow the prediction algorithm
to output anything it wants and define $t_{f}$ by counting only the
valid predictions in the segment that immediately follows $t=T$.
At this point, there seems to be little to choose between the two
possibilities. So we adopt the more restrictive definition of a prediction
algorithm.

Now our claim can be stated as follows: if $P$ is a valid prediction
algorithm
\begin{equation}
\limsup_{T\rightarrow\infty}\frac{t_{f}}{\log_{2}T}<\frac{1+\epsilon}{H}\label{eq:prediction-bound}
\end{equation}
with probability $1$ for any $\epsilon>0$. The notion of entropy
$H$ that we adopted in the previous section was metric entropy $h_{\mu}(f)$
relative to the physical measure $\mu$ on $\mathcal{A}$. For a hyperbolic
attractor, the physical measure is the SRB measure and it is guaranteed
to exist. Thus we are assuming $X_{0}$ to be $\mu$-distributed,
$X_{1}=f\left(X_{0}\right)$, $X_{2}=f\left(X_{1}\right)$, and so
on. 

In order to explain why every prediction algorithm must satisfy the
bound \prettyref{eq:prediction-bound}, we turn to another notion
of entropy, namely topological entropy $h_{top}(f)$ \cite{KatokHasselblatt1997}.
To begin with we have a metric $d$ on $\mathcal{A}$. We can define
$d_{n}(x,y)$ to be the maximum of $d\left(f^{i}(x),f^{i}(y)\right)$
over $i=0,\ldots,n-1$. If $N(\delta,n)$ is the number of $\delta$
balls required to cover $\mathcal{A}$ in the metric $d_{n}$, topological
entropy is defined using the relation $N(\delta,n)\approx C2^{nh_{top}}$
for small $\delta$. It is independent of the metric. In general,
$h_{\mu}(f)\leq h_{top}(f)$ (see Theorem 4.5.3 of \cite{KatokHasselblatt1997}).
With the assumptions we have made on $\mathcal{A}$ and $\mu$, $h_{\mu}=h_{top}$.

Suppose we are given the sequence $X_{0},\ldots,X_{T}$. That is equivalent
to assuming that we know the iterates $f$, $f^{2}$,$\ldots$, $f^{n}$
at $T-n+1$ points on $\mathcal{A}$. For example, we know $f(X_{1})=X_{2}$,
$f^{3}(X_{2})=X_{5}$, and so on. We are assuming $n$ to be of the
order of $\log_{2}T$. These $T-n+1$ points on $\mathcal{A}$ at
which $n$ iterates of $f$ are known may be assumed to be approximately
$\mu$-distributed \cite{Young2002}. To predict $n$ iterates of
$X_{T}$ with tolerance $tol=\delta$ from that information, we require
one of the points $X_{0},\ldots,X_{T-n}$ to be within $\delta$ of
$X_{T}$ in the $d_{n+1}$ metric. For such a thing to be possible,
we require $T-n\geq C2^{nh_{top}}$ or $n\leq\log_{2}T/h_{top}$ asymptotically.

It may seem that one may extract some more information about $f^{n}\left(X_{T}\right)$
by clever interpolation of $f^{n}$ whose value is known at $X_{0},\ldots,X_{T-n}$.
It is true that clever interpolation can improve the accuracy dramatically
if the function being interpolated is smooth. In this context, however,
no such thing is possible even if $f$ is infinitely differentiable
or real analytic. The key reason is that the exponential divergence
of trajectories is enough to defeat any attempt at clever interpolation.

Perhaps this point will be clearer with an example. The map $x_{n+1}=f(x_{n})$
with $f(x)=4x(1-x)$ over the interval $[0,1]$ has topological entropy
equal to $1$. Suppose we want to predict $f^{n}\left(x_{T}\right)$.
Given the shape of $f$, $f^{n}$ will have $2^{n-1}$ oscillations.
By an oscillation we mean a monotonic increase in $f^{n}(x)$ from
$0$ to $1$ and then a monotonic decrease to $0$. If $T<C2^{nh_{top}/(1+\epsilon)}=C2^{n/(1+\epsilon)}$,
it is clear that $T$ points will be too few to track all the oscillations
of $f^{n}$. No interpolation scheme can make up for that kind of
undersampling. 

As indicated earlier, the theoretical discussion in this section is
restricted to maps. However, a new point comes up in relation to flows
that is worth mentioning. Suppose we have a continuous signal $x(t)$
for $0\leq t\leq T$ from a real analytic flow. Then $x(t)$ is analytic
in a neighborhood of the real line. Thus in principle we may use the
known stretch of the signal to predict it forever into the future
using analytic continuation. Analytic continuation is numerically
unstable and often not feasible as an extrapolation strategy. Limitations
to the applicability of analytic continuation become evident the moment
we note that the continuous signal must be sampled at some finite
rate and that it is incorrect to assume the entire signal to be available.
A very similar point comes up in the context of the Wiener-Kolmogorov
predictor. See Section 1.7 of \cite{Wiener1949}.

A prediction algorithm $P$ is optimal if 
\begin{equation}
\liminf_{T\rightarrow\infty}\frac{t_{f}}{\log_{2}T}\geq\frac{1-\epsilon}{H}\label{eq:optimal-prediction}
\end{equation}
with probability $1$ for any $\epsilon>0$. Our view of optimality
is tied to almost sure prediction and not to statistical predictability.
The practical significance of almost sure convergence is accepted
in information theory. See the discussion in \cite{WynerZiv1989}
for an example.

\section{The embedding predictor, related predictors, and their suboptimality}

With regard to Wiener-Kolmogorov predictors, Wiener wrote \cite[p. 71]{Wiener1949}:
\emph{geometrical facts must be predicted geometrically and analytical
facts analytically, leaving only statistical facts to be predicted
statistically}. 

There are two geometrical facts that are central to the prediction
of chaotic signals. The first is recurrence and the second is the
need to decompose close recurrences into stable and unstable components.
Existing predictors have exploited recurrence but have not attempted
to decompose close recurrences into stable and unstable components.
As a result, they fall well short of being optimal in the sense of
\prettyref{eq:optimal-prediction}.

In this section, we discuss a few existing predictors of chaotic signals.
Some of the ideas used by existing predictors, which we find to be
deficient with respect to optimal prediction, may become more useful
once a good method is found to decompose close recurrences into stable
and unstable components. For example, polynomial interpolation has
been suggested and used for limited improvement of the accuracy of
predictions of chaotic time series. It is of little use in getting
closer to optimality. However, if close recurrences are decomposed
appropriately into stable and unstable components, polynomial interpolation
may indeed be useful for improving the accuracy of the prediction
of $x(T+s)$, especially for $s<\alpha\log_{2}T/H$, where $\alpha$
is a small fraction.

\begin{figure}
\begin{centering}
\includegraphics[scale=0.6]{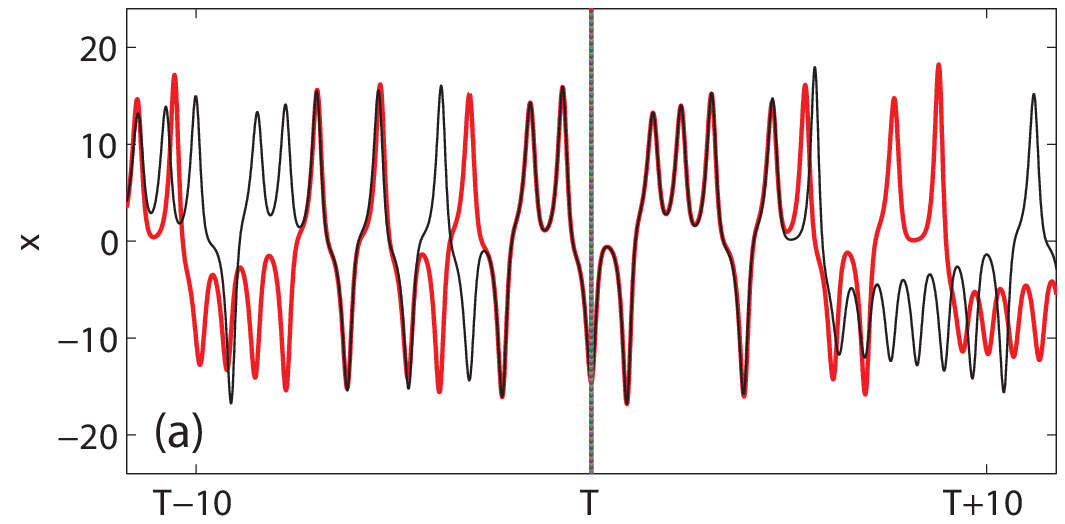}
\par\end{centering}

\begin{centering}
\includegraphics[scale=0.6]{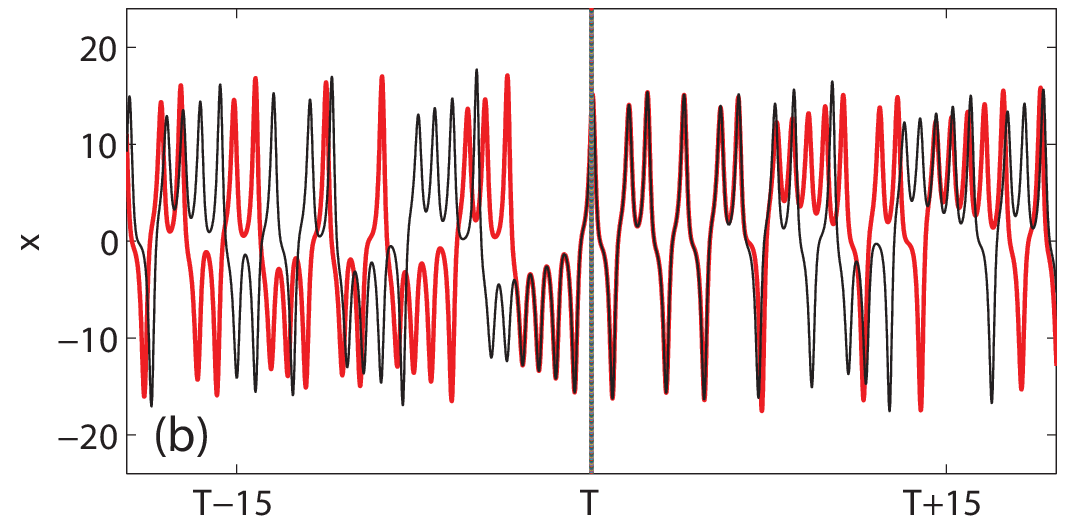}
\par\end{centering}

\centering{}\caption{Suboptimal predictions (in thick red) using the embedding predictor
to a Lorenz signal (in thin black). (a) $T=2^{14}$ symbols. (b) $T=2^{21}$
symbols. \label{fig:Lorenz-embedding}}
\end{figure}

Phase space reconstruction using delay coordinates is used by all
existing predictors. We term the most basic of these predictors as
the embedding predictor \cite{Casdagli1989,FarmerSidorowich1987}.
Given a signal $x(t)$ for $0\leq t\leq T$, the embedding predictor
finds $t^{\ast}$ to minimize \prettyref{eq:embed-fit}, as we have
already discussed. The key idea behind embedding predictors is to
indirectly recover the location of the dynamical system in phase space
at time $t$ using the delay coordinates $\left(x(t),x(t-\tau),\ldots,x(t-(k-1)\tau)\right)$.
Suppose the state vector of the dynamical system at time $t=t_{1}$
is $X_{1}$ and the state vector at time $t=t_{2}$ is $X_{2}$. It
is quite possible that $x(t_{1})=x(t_{2})$ even if $X_{1}\neq X_{2}$
or that $\bigl|x\left(t_{1}\right)-x\left(t_{2}\right)\bigr|$ is
small even if $X_{1}$ is not close to $X_{2}$. However, the pattern
of events preceding $t=t_{1}$ and $t=t_{2}$ as recorded using delay
coordinates gives us better information to decide if $X_{1}$ and
$X_{2}$ are close to each other or not.

Although the choice of the delay parameter $\tau$ and the embedding
dimension $k$ have been discussed extensively, it is difficult to
make definite statements about what the best choices are. One approach
is to use mutual information---see \cite{Abarbanel1996}. In this
approach it is assumed that $\tau$ should not be too small because
nearby values are well-correlated and not too large because distant
points on the signal are very weakly correlated. Mutual information
is used to find some kind of a compromise. Regarding the embedding
dimension $k$, it is stated that it should be at least as large as
the dimension of the underlying chaotic set. For the validity of Taken's
embedding theorem \cite{Takens1981}, $k\geq2m+1$ is required with
$m$ being the dimension of the chaotic set.

Figure \ref{fig:Lorenz-embedding} shows suboptimal predictions of
Lorenz signals using the embedding predictor and $\tau=.03$, $k=5$.
Our choice of the delay parameter at $\tau=.03$ is much smaller than
what the mutual information criterion would imply. The mutual information
criterion would imply a $\tau$ that is large enough to span a few
oscillations of the signal. It is difficult to see what advantage
using information from such distant points may have with regard to
prediction, where the game is to exploit local information optimally.
Indeed, use of a larger delay parameter gives no improvement at all.
Some of the extant discussion about choosing the delay parameter appears
to be based on a desire to obtain good plots and not good predictions. 

For a study of the effect of the delay parameter on the quality of
prediction, see Figure 22 of Casdagli et al.\cite{CasdagliGibson1991}.
For the Ikeda map, the optimal delay for prediction is found to be
the smallest delay possible. In Figure 22 of that paper, an attempt
is made to predict only one iteration using a history that is equal
to $10^{4}$ iterates in length. The entropy theorem indicates that
more than 20 iterates could be predictable using a history of that
length. Here the advantage of defining optimal prediction as in \prettyref{eq:optimal-prediction},
which we mentioned earlier in the introduction, becomes evident in
a more concrete way. If we attempt to predict only one iterate, different
prediction methods will differ in terms of accuracy, but the difference
will be quite delicate. Even for predicting a single iterate optimally,
it is important to resolve close recurrences into stable and unstable
components. However, the gain in accuracy to be obtained by resolving
close recurrences in that manner is not easily noticed. In contrast,
the optimality criterion \prettyref{eq:optimal-prediction} which
emphasizes the length of the fit into the future, exposes the central
deficiency of existing predictors in a way that is quite easy to see.

If we compare Figure \ref{fig:Lorenz-embedding} with Figure \ref{fig:Lorenz-best-fit},
it is abundantly clear that the embedding predictor does not extract
the information in the history of the signal in an optimal manner.
The embedding predictor gives a closer fit in the immediate past of
$t=T$, but that is precisely why it does not do the best job of predicting
the future. Still from Figure \ref{fig:Lorenz-embedding}, we see
that the fit into the future is much better than the fit into the
past. Does the embedding method have a bias to the future after all?
The answer is no. The embedding method treats the past and the future
equally. There is nothing in it to say that it is attempting to predict
the future rather than fit the past. The better fit into the future
we see in the figure is a consequence of the Lyapunov exponents of
the Lorenz attractor. The lone negative exponent of the Lorenz attractor
is $-14.5$ (using natural logarithms) and is much larger in magnitude
that the lone positive exponent, which is $0.905$. Therefore if we
pick two points close to each other on the Lorenz attractor, the corresponding
trajectories will typically diverge faster in the past. 

{\footnotesize }
\begin{table}
\begin{centering}
{\footnotesize }%
\begin{tabular}{|>{\centering}p{0.75in}|>{\centering}p{0.75in}||>{\centering}p{0.75in}|>{\centering}p{0.75in}|}
\hline 
{\footnotesize $\log_{2}T$ }{\footnotesize \par}

{\footnotesize (in symbols)} & {\footnotesize $t_{embed}$}{\footnotesize \par}

{\footnotesize (in symbols)} & {\footnotesize $\log_{2}T$ }{\footnotesize \par}

{\footnotesize (in symbols)} & {\footnotesize $t_{embed}$}{\footnotesize \par}

{\footnotesize (in symbols)}\tabularnewline
\hline 
\hline 
{\footnotesize 2} & {\footnotesize 0} & {\footnotesize 12} & {\footnotesize 8}\tabularnewline
\hline 
{\footnotesize 3} & {\footnotesize 0} & {\footnotesize 13} & {\footnotesize 10}\tabularnewline
\hline 
{\footnotesize 4} & {\footnotesize 1} & {\footnotesize 14} & {\footnotesize 7}\tabularnewline
\hline 
{\footnotesize 5} & {\footnotesize 5} & {\footnotesize 15} & {\footnotesize 9}\tabularnewline
\hline 
{\footnotesize 6} & {\footnotesize 6} & {\footnotesize 16} & {\footnotesize 7}\tabularnewline
\hline 
{\footnotesize 7} & {\footnotesize 4} & {\footnotesize 17} & {\footnotesize 9}\tabularnewline
\hline 
{\footnotesize 8} & {\footnotesize 4} & {\footnotesize 18} & {\footnotesize 10}\tabularnewline
\hline 
{\footnotesize 9} & {\footnotesize 6} & {\footnotesize 19} & {\footnotesize 9}\tabularnewline
\hline 
{\footnotesize 10} & {\footnotesize 2} & {\footnotesize 20} & {\footnotesize 8}\tabularnewline
\hline 
{\footnotesize 11} & {\footnotesize 3} & {\footnotesize 21} & {\footnotesize 9}\tabularnewline
\hline 
\end{tabular}
\par\end{centering}{\footnotesize \par}

{\footnotesize \caption{Length of suboptimal predictions of a Lorenz signal ($t_{embed}$)
using the embedding predictor.\label{tab:Lorenz-embed-fwd}}
}
\end{table}
The $t_{embed}$ column of Table \ref{tab:Lorenz-embed-fwd} is obtained
as follows. The metric \prettyref{eq:embed-fit} is used to pick $t^{\ast}$
so that the distance between the delay coordinates at $t=t^{\ast}$
and $t=T$ is the smallest. The length of the fit into the future
is given by $t_{embed}$: $|x\left(t^{\ast}+s\right)-x(T+s)|\leq tol$
for $0\leq s\leq t_{embed}$ but not for $0\leq s\leq t$ with $t>t_{embed}$.
Comparison of $t_{embed}$ in Table \ref{tab:Lorenz-embed-fwd} with
$t_{best}$ in Table \ref{tab:Lorenz-best-fit} shows that the embedding
predictor does not approach optimality.

In the rest of this section, we consider a number of extant ideas
for improving the basic embedding predictor. All these ideas have
merits. However, to be fully effective, they need to take into account
an essential aspect of chaotic signals, which is their tendency to
separate or come together depending upon the relative sizes of the
stable and unstable components. 

The first idea we mention is from the paper by Farmer and Sidorowich
\cite{FarmerSidorowich1987}. To predict $x(T+s)$ the basic embedding
predictor picks a single $t^{*}\in[k\tau,T-s]$ using the metric \prettyref{eq:embed-fit}.
Instead, a predictor may pick $p$ different instants $t_{1}^{\ast},\ldots,t_{p}^{\ast}$
where the delay coordinates are the $p$ closest to the delay coordinates
at $t=T$. Assuming $p\geq k$, the prediction of $x(T+s)$ is generated
as a linear combination of the delay coordinates at $t=T$ by fitting
$x(t_{i}^{\ast}+s)$ as a linear combination of the delay coordinates
at $t=t_{i}^{\ast}$, for $i=1,\ldots,p$, using linear least squares.

Let us first understand the merit of this idea. Suppose we are looking
at a Lorenz signal and we fix $s=1$, which means we are trying to
predict the signal at a point that is somewhat more than one return
time ($t_{return}=0.7511$) from $t=T$. For sufficiently large $T$,
the signal will have delay coordinates at $t=t_{i}^{\ast}$ close
to that at $t=T$ for each of the $p$ values of $i$. More importantly,
they will be sufficiently close that none of the $p$ segments $x(t),\, t_{i}^{\ast}\leq t\leq t_{i}^{\ast}+s$,
will diverge from each other for $i=1,\ldots,p$. Therefore extrapolation
using least squares will improve the order of accuracy (see Figure
2 of \cite{FarmerSidorowich1987}).

The situation is quite different if we take $s=\alpha\log_{2}T/H$,
with say $\alpha=0.75$. In this case, we want to predict an instant
that gets farther out in time as $T$ increases. In this situation
the $p$ segments $x(t),\, t_{i}^{\ast}\leq t\leq t_{i}^{\ast}+s,$
with $i=1,\ldots,p$ will diverge from each other with high probability
ruining any attempt to extrapolate using linear least squares. One
may attempt to patch the situation by trying to classify the $p$
segments into clusters that stay close to each other and then picking
one of the clusters to extrapolate from $t=T$ to $t=T+s$. But to
do so would be to get back to our point that one has to decompose
the distance between segments of the signal into stable and unstable
components for optimal prediction.

Even with $s=1$, in which case extrapolation using least squares
improves the accuracy of the basic embedding predictor, there are
advantages to decomposing the distance between segments of the signal
into stable and unstable components. Such a decomposition will allow
us to weight the different segments from the past and wring all the
information out of the signal. Conversely, ideas such as extrapolation
using linear least squares may be useful once the basic issue of resolving
the distance between segments into stable and unstable components
is addressed. 

Other ideas for improving the basic embedding predictor are to use
higher order polynomials for extrapolation \cite{FarmerSidorowich1987},
to trap the delay coordinates at $t=T$ within a simplex in reconstructed
phase space \cite{SugiharaMay1990}, or to weight close recurrences
using the closeness of the approach \cite{Abarbanel1996}. The merits
and demerits of these ideas are as in the discussion above and nothing
more needs to be said. Another idea is to extrapolate from $t=T$
to $t=T+1$ using the embedding predictor possibly with enhancements
and then iterate the extrapolation from $t=T$ to $t=T+1$ a total
of $s$ times to extrapolate from $t=T$ to $t=T+s$. The merit of
this idea is to bring in new information from the signal to evaluate
intermediate points such as $t=T+1$ and $t=T+2$. However, the embedding
predictor continues to be suboptimal even with this enhancement. The
problem is that a single step of extrapolation will throw away all
the information about stable and unstable manifolds in the vicinity
of $t=T$ . The way the stable and unstable components of the distance
between two segments of the signal must be taken into account depends
upon how far into the future we want to extrapolate, as will become
clear in the next section.

\section{Character of an optimal predictor}

In this section, we give a sense of how an optimal predictor might
work. Although a general purpose optimal predictor has not yet been
derived, it is possible to give a sense of what such a predictor should
do. 

Suppose $c$ is a fixed point of the map $f$. The iterates at $c$
will obviously look like
\[
c,c,c,\ldots
\]
Suppose we pick a point $X_{0}$ within a distance $\epsilon$ of
$c$ and look at the sequence 
\[
X_{0},f\left(X_{0}\right),f^{2}\left(X_{0}\right),\ldots
\]
When is the latter sequence closest to the former sequence? The answer
is they are closest when $X_{0}$ lies on the stable manifold of $c$.
If it lies on the unstable manifold of $c$, on the other hand, the
latter sequence will quickly diverge from the former. Here we already
see the basic ingredient for optimal prediction. For a good match
between the sequences, it is not enough to pick $X_{0}$ close to
$c$ but we have to pick $X_{0}$ to be on or close to the stable
manifold of $c$. An optimal predictor has to implement this idea
using time series data and nothing more.

In general, it is impossible to pick a point that is exactly on the
stable manifold. Therefore, we expand upon what it means to pick a
point that is close to the stable manifold. Let $c$ be a point on
the hyperbolic attractor. Let us suppose that $x$ is close enough
to $c$ and that we may write $x$ as 
\begin{equation}
x=c+\sum_{i=1}^{u}a_{i}v_{i}^{+}(c)+\sum_{i=1}^{s}b_{i}v_{i}^{-}(c).\label{eq:xc}
\end{equation}
Here $v_{i}^{+}(x)$ are unit vectors in the tangent space at $x$
corresponding to positive Lyapunov exponents and the $v_{i}^{-}(x)$
are unit vectors corresponding to negative Lyapunov exponents. For
simplicity, we assume the Lyapunov exponents to be distinct with $u$
positive exponents and $s$ negative exponents. Let $\lambda_{i}^{+}$
be the characteristic multiplier corresponding to $v_{i}^{+}$ and
similarly let $\lambda_{i}^{-}$ correspond to $v_{i}^{-}$ (if $l$
is a Lyapunov exponent defined using natural logarithms, $\exp(l)$
is the corresponding characteristic multiplier). We have 
\begin{equation}
f^{n}(x)\approx c_{n}+\sum_{i=1}^{u}a_{i}\left(\lambda_{i}^{+}\right)^{n}v_{i}^{+}(c_{n})+\sum_{i=1}^{s}b_{i}\left(\lambda_{i}^{-}\right)^{n}v_{i}^{-}(c_{n})\quad\text{where}\quad c_{n}=f^{n}(c).\label{eq:fnx-cn}
\end{equation}
Here we have assumed that the expansion along the directions $v_{i}^{+}$
and $v_{i}^{-}$ is by the same factor with each iteration. With that
assumption, it is easier to bring out the essential aspects of the
heuristic argument we are developing here. Note that $\abs{\lambda_{i}^{+}}>1$
and $\abs{\lambda_{i}^{-}}<1$. 

To eliminate some linear algebra from the discussion, we will assume
that $v_{i}^{+}(x)$, $1\leq i\leq u$, and $v_{i}^{-}(x)$, $1\leq i\leq s$,
form an orthonormal basis for the tangent space at each point $x$
on the hyperbolic attractor. For the related concepts of adapted metric
and adapted coordinates, see \cite{KatokHasselblatt1997}.

Suppose (as usual) that the points in the available trajectory are
$x_{0},\ldots,x_{T}$ with $x_{T}=c$. To predict the sequence $f(c),f^{2}(c),\ldots$
,$f^{k}(c)$, with $k\approx\log_{2}T/H$, we will look at points
from the sequence $x_{0},\ldots,x_{T-k}$ that are close enough to
$x_{T}=c$ and can be represented in the form \prettyref{eq:xc}.
Here we will examine what kind of points $x$ are available in the
sequence and which ones will be useful predictors.

Let us try to find an $x$ of the form \prettyref{eq:xc} in the available
history with $a_{i}=A_{i}\delta$ for $1\leq i\leq u$ and $b_{i}=B_{i}\delta$
for $1\leq i\leq s$ with $A_{i}$ and $B_{i}$ fixed to determine
the shape of the box around $c$ and with as small a $\delta$ as
possible. Kac's theorem (Theorem \ref{thm:Kac}) suggests that we
may find a point in the available history in a box around $c$ if
the volume of the box is $1/T$ or more. Thus in a box of shape determined
by $A_{i}$ and $B_{i}$, the smallest $\delta$ that leaves the box
large enough to be likely to include a point from the available history
is given by $A_{1}\ldots A_{u}B_{1}\ldots B_{s}\delta^{s+u}\approx1/T$.
In fact, we will allow the stable components $b_{i}$ to be as large
as the tolerance allows. In that case, the box has dimensions $a_{i}=A_{i}\delta$
and $b_{i}=O(1)$. The smallest delta should then satisfy 
\begin{equation}
A_{1}\ldots A_{u}\delta^{u}\approx\frac{C}{T}\label{eq:A1..Audeltau}
\end{equation}
for some constant $C$.

We may now try to choose the shape of the box to allow $f^{n}(x)$
to stay close to $f^{n}(c)$ for $n=1,\ldots,k$. If we estimate the
distance between $f^{n}(x)$ and $f^{n}(c)$ using \prettyref{eq:fnx-cn},
the distance comes out as follows:
\begin{equation}
\norm{f^{n}(x)-f^{n}(c)}\approx\sqrt{\sum_{i=1}^{u}a_{i}^{2}\left(\lambda_{i}^{+}\right)^{2n}}.\label{eq:norm{fn(x)-fn(c)}approx}
\end{equation}
Here we have neglected the $\mu_{i}^{-}$ components because $\abs{\mu_{i}^{-}}<1$
and these stable components diminish rapidly with $n$. As long as
the stable components are less than a tolerance, we do not need to
worry about them. Given the constraint on how small the box can get,
the best shape is obtained by taking $A_{i}=1/\left(\lambda_{i}^{+}\right)^{n}$.
The value of $\delta$ implied by \prettyref{eq:norm{fn(x)-fn(c)}approx}
is 
\begin{equation}
\delta^{u}\approx\frac{C\left(\prod\lambda_{i}^{+}\right)^{n}}{T}\label{eq:deltauapprox}
\end{equation}
and the minimum possible value of $\norm{f^{n}(x)-f^{n}(c)}$ is approximately
$\delta\sqrt{u}$.

From this heuristic calculation, we learn two things. If we want to
pick an $x$ from the available history to minimize $\norm{f^{n}(x)-f^{n}(c)}$
it is not enough to simply pick an $x$ from the history that is as
close to $c$ as possible. We have to balance the sizes of the unstable
components $a_{i}$ carefully. The stable components $b_{i}$ can
be as large as the tolerance of the problem allows, which means that
the best $x$ for predicting $f^{n}(c)$ may not be particularly close
to $c$. 

For valid prediction of $f^{n}(c)$, we require $\norm{f^{n}(x)-f^{n}(c)}\approx\delta\sqrt{u}\leq tol$.
If we use expression \prettyref{eq:deltauapprox} for $\delta$, we
get 
\begin{equation}
n\leq\frac{\log_{2}T+u\log_{2}tol-(u/2)\log_{2}u-\log_{2}C}{\sum\log_{2}\lambda_{i}^{+}}.\label{eq:nleqlogTbysumofloglambda}
\end{equation}
For a hyperbolic attractor, metric entropy is equal to $\sum\log_{2}\lambda_{i}^{+}$.
From this calculation, we understand why the metric entropy shows
up the way it does in the entropy theorem.

In the argument leading up to \prettyref{eq:nleqlogTbysumofloglambda}
, we assumed $x$ and $c$ to be points on the hyperbolic attractor.
A predictor which predicts $x_{T+n}$ for $n$ that approaches the
upper bound in \prettyref{eq:nleqlogTbysumofloglambda} or is optimal
in the sense of \prettyref{eq:optimal-prediction} has to calculate
the $a_{i}$ in \prettyref{eq:xc} using time series data alone.

Given a Lorenz signal, suppose we want to assess if $t=t^{\ast}$
will give a long fit to the segment following $x(T)$, with the length
of fit defined as in \prettyref{eq:length-of-fit} . If we knew the
points $X(t^{\ast})$ and $X(T)$ in the three-dimensional phase space
of the Lorenz flow, as well as the decomposition $X(t^{\ast})-X(T)=s+f+u$---where
$s$ is along the stable direction at $X(T)$, $f$ is along the flow
at $X(T)$, and $u$ is along the unstable direction at the same point---the
assessment would be easy to make. As long as the components $f$ and
$s$ are below the tolerance, we want the minimum $\norm{u}$ possible
for the longest fit. 

The embedding method attempts to estimate the distance between $X(t^{\ast})$
and $X(T)$ using delay coordinates and the formula \prettyref{eq:embed-fit}.
It does not even attempt to resolve the close recurrences into $s$,
$f$, and $u$ components as an optimal predictor should.

\section{Optimal prediction of toral automorphisms}

Let $A$ be a $d\times d$ matrix with integer entries and $\det A=\pm1$.
The map $X_{n+1}=AX_{n}\bmod1$ is a hyperbolic toral automorphism
if no eigenvalue of $A$ has unit modulus. Here $X_{n}$ is a vector
with $d$ entries each of which is assumed to be in the interval $[0,1)$.
Each entry of the matrix vector product $AX_{n}$ is taken modulo
$1$ in the interval $[0,1)$ to get $X_{n+1}$. The space $[0,1)^{d}$
is used as the coordinate space of the torus $\mathbb{T}^{d}$. 

The class of hyperbolic toral automorphisms is a basic example in
theoretical dynamics \cite{KatokHasselblatt1997}. Such automorphisms
are topologically transitive on the torus and possess Markov partitions
of arbitrarily small diameter. The physical measure is the Lebesgue
measure and the entropy is positive.

We will consider the prediction of the signal $x_{0},\ldots,x_{T}$,
where $x_{n}$ is the first entry of $X_{n}$ for each $n$, $X_{0}$
is uniformly distributed on $\mathbb{T}^{d}$, and $X_{n+1}=AX_{n}\bmod1$
for $n\ge0$. The first toral automorphism that is considered is 
\begin{equation}
X_{n+1}=AX_{n}\bmod1,\quad A=\left(\begin{array}{cc}
2 & 1\\
1 & 1
\end{array}\right).\label{eq:toral-auto-2d}
\end{equation}
This matrix $A$ has eigenvalues $\frac{3+\sqrt{5}}{2}\approx2.61803$
and $\frac{3-\sqrt{5}}{2}\approx0.381966$ and its entropy is $\log_{2}2.61803=1.3885$.
The second toral automorphism that is considered is 
\begin{equation}
X_{n+1}=AX_{n}\bmod1,\quad A=\left(\begin{array}{ccc}
0 & -1 & 0\\
1 & -2 & 1\\
2 & -3 & 3
\end{array}\right).\label{eq:toral-auto-3d}
\end{equation}
This matrix A has eigenvalues $2.1479$ and $-0.57395\pm i0.368989$.
In both instances, $\det A=1$.

Before considering the optimal prediction of signals derived from
toral automorphisms, it is important to note that restricting ourselves
to the class of hyperbolic toral automorphisms means that some oddities
occur that would not occur with a general purpose optimal predictor.
Hyperbolic toral automorphism of dimension $d$ are defined using
finitely many parameters each of which is an integer (entries of the
matrix $A$). One may exploit that fact and tweak the predictor in
the next section to reconstruct the toral automorphism exactly. We
do not overly specialize the prediction scheme in that way. The purpose
of the prediction scheme is to show what kind of considerations may
arise in the derivation of a general purpose predictor and the exact
reconstruction of the toral automorphism from time series data is
irrelevant in that regard. 

In the previous section, we have emphasized that close recurrences
must be resolved into stable and unstable components and the quantities
$a_{i}(\lambda_{i}^{+})$ that appear in \prettyref{eq:fnx-cn} must
be estimated. For hyperbolic toral automorphisms, the stable and unstable
directions split in exactly the same way at every point on the torus.
The optimal predictor based on Pade approximation that we derive takes
advantage of this fact and limits itself to estimating $a_{i}$. The
difficulty of estimating stable and unstable directions near close
recurrences, which must be tackled by a general purpose predictor,
are sidestepped by the Padé predictor.

We begin by considering the so-called exponential extrapolation problem.
Suppose a sequence is defined by 
\begin{equation}
s_{n}=\sum_{k=1}^{d}c_{k}\lambda_{k}^{n}\quad n=0,1,\ldots\label{eq:sn-sumofexps}
\end{equation}
The problem is to find $s_{2d}$, $s_{2d+1}$, and so on given $s_{0},\ldots,s_{2d-1}$.
Since the sequence is defined by $d$ parameters $c_{k}$ and $d$
parameters $\lambda_{k}$, it is reasonable to expect that the first
$2d$ numbers of the sequence may determine the rest of the sequence.
The exponential extrapolation problem is to determine the rest of
the sequence. It was solved by Prony late in the 18th century (see
\cite{Hamming1986} for a discussion of Prony's method). We present
a solution based on Padé approximants. Our presentation could be new.
Padé approximants generalize naturally to vector Padé approximants,
which may turn out to be useful in deriving a general purpose predictor.
For an introduction to Padé approximation, see \cite{Baker1975}.

Define $f(z)=\sum_{n=0}^{\infty}s_{n}z^{n}$. Using \prettyref{eq:sn-sumofexps},
we get 
\[
f(z)=\sum_{k=1}^{d}\frac{c_{k}}{1-\lambda_{k}z}=:\frac{\alpha_{0}+\alpha_{1}z+\cdots+\alpha_{d-1}z^{d-1}}{1+\beta_{1}z+\cdots+\beta_{d}z^{d}}.
\]
The right hand side is the $(d-1,d)$ Padé approximant of $f(z)$.
Determining the $\beta_{i}$ is the key to exponential extrapolation.
We have 
\begin{eqnarray*}
\alpha_{0}+\cdots+\alpha_{d-1}z^{d-1} & = & \left(1+\beta_{1}z+\cdots+\beta_{d}z^{d}\right)\sum_{n=0}^{\infty}s_{n}z^{n}\\
 & = & \sum_{n=0}^{\infty}z^{n}\left(s_{n}+\sum_{j=1}^{\min(d,n)}s_{n-j}\beta_{j}\right).
\end{eqnarray*}
Equating coefficients of $z^{n}$ for $n=d,\ldots,2d-1$, we get the
$d$ equations
\begin{equation}
\sum_{j=1}^{d}s_{n-j}\beta_{j}=-s_{n}.\label{eq:sk-sum}
\end{equation}
This Toeplitz system must be solved to determine $\beta_{j}$. Its
solvability is a necessary condition for exponential extrapolation.
Once the $\beta_{j}$ are determined, \prettyref{eq:sk-sum} is used
with $n=2d,2d+1,\ldots$ to determine $s_{2d}$, $s_{2d+1}$, and
so on.

The analogy of this process to the Wiener-Kolmogorov predictor described
in \cite{Wiener1949} is unmistakable. In both cases, a Toeplitz system
must be solved. Once the Toeplitz system is solved, new numbers in
the sequence are obtained as fixed linear combinations of preceding
numbers in the sequence. Indeed, it is quite possible that there may
be a way to view the Wiener-Kolmogorov predictors as variations or
extensions of Prony's method as presented here. The Toeplitz system
that comes up in exponential extrapolation is unsymmetric in general,
while the Toeplitz system that comes up in the Wiener-Kolmogorov predictor
is symmetric. 

\begin{table}
\begin{centering}
\begin{tabular}{|c|c|c|c||c|c|c|c|}
\hline 
$\log_{2}T$ & $t_{best}$ & $t_{embed}$ & $t_{pade}$ & $\log_{2}T$ & $t_{best}$ & $t_{embed}$ & $t_{pade}$\tabularnewline
\hline 
\hline 
5 & 2 & 1 & 1 & 17 & 12 & 4 & 8\tabularnewline
\hline 
6 & 7 & 0 & 0 & 18 & 10 & 4 & 10\tabularnewline
\hline 
7 & 3 & 0 & 1 & 19 & 13 & 6 & 10\tabularnewline
\hline 
8 & 5 & 1 & 0 & 20 & 15 & 6 & 11\tabularnewline
\hline 
9 & 4 & 2 & 2 & 21 & 15 & 6 & 13\tabularnewline
\hline 
10 & 4 & 2 & 2 & 22 & 14 & 1 & 12\tabularnewline
\hline 
11 & 8 & 2 & 3 & 23 & 19 & 6 & 19\tabularnewline
\hline 
12 & 6 & 2 & 5 & 24 & 15 & 9 & 13\tabularnewline
\hline 
13 & 13 & 1 & 5 & 25 & 17 & 7 & 17\tabularnewline
\hline 
14 & 8 & 6 & 6 & 26 & 16 & 9 & 15\tabularnewline
\hline 
15 & 8 & 4 & 6 & 27 & 18 & 7 & 18\tabularnewline
\hline 
16 & 11 & 0 & 6 & 28 & 18 & 1 & 16\tabularnewline
\hline 
\end{tabular}
\par\end{centering}

\caption{Length of best fit from the past, suboptimal prediction using the
method of embedding, and optimal Padé prediction of a signal obtained
from the automorphism \prettyref{eq:toral-auto-2d} of the two dimensional
torus $\mathbb{T}^{2}$.\label{tab:torus-2d}}
\end{table}

Let $x_{0},\ldots,x_{T}$ be a signal obtained from a hyperbolic toral
automorphism as explained in the previous section. Suppose we want
to compare the segment
\[
x_{t^{\ast}-2d+1},\ldots,x_{t^{\ast}-1},x_{t^{\ast}}
\]
with the segment 
\[
x_{T-2d+1},\ldots,x_{T-1},x_{T}.
\]
We first form the differences $\Delta x_{i}=x_{t^{\ast}-2d+1+i}-x_{T-2d+1+i}$
for $i=0,\ldots,2d-1$. Our intention is to extrapolate the $\Delta x_{i}$
sequence to figure out how well $x_{t^{\ast}+s}$ will predict $x_{T+s}$.
Since the toral automorphisms are carried out modulo $1$, we begin
by making the following modification to the $\Delta x_{i}$ sequence.
For each $i$ with $0\leq i\leq2d-1$, if $\Delta x_{i}>1/2$, we
replace $\Delta x_{i}$ by $\Delta x_{i}-1$. On the other hand, if
$\Delta x_{i}\leq-1/2$, we replace $\Delta x_{i}$ by $\Delta x_{i}+1$.
After these operations, we will have $\abs{\Delta x_{i}}\leq1/2$
for $i=0,\ldots,2d-1$. 

If the point on the torus $\mathbb{T}^{d}$ that corresponds to $x_{n}$
is $X_{n}$, we have $X_{n+1}-X_{m+1}=A(X_{n}-X_{m})\bmod1$. Therefore
if $X_{t^{\ast}-2d-1}-X_{T-2d-1}$ is small enough, the sequence $\Delta x_{i}$,
$i=0,\ldots,2d-1$, can we written as a linear combination of exponentials
like the $s_{i}$ sequence in \prettyref{eq:sn-sumofexps}. The $\lambda_{i}$
will be the eigenvalues of $A$. We use a tolerance to check if the
$\Delta x_{i}$ are small enough to permit sensible exponential extrapolation.

Using exponential extrapolation, we compute $\Delta x_{2d}$, $\Delta x_{2d+1}$,
and so on, and find the maximum $n$ such that each of the numbers
\[
\abs{\Delta x_{2d}},\ldots,\abs{\Delta x_{2d+n-1}}
\]
is less than $tol$. For the computations reported in this section,
$tol=0.1$. The $n$ found in this way is the expected length of fit.
The $t^{\ast}$ which gives the maximum expected length of fit is
chosen. The sequences $x_{t*+1},x_{t^{\ast}+2},\ldots$ and $x_{T+1},x_{T+2},\ldots$
are compared to determine the actual length of fit, which is denoted
by $t_{pade}$.

In Table \ref{tab:torus-2d}, we list $t_{best}$ (the best fit from
the past defined as in \prettyref{eq:length-of-fit}), $t_{embed}$,
and $t_{pade}$. For the embedding predictor, we took $2d$ to be
the embedding dimension. By going down the table, we can easily detect
that the entropy $H$ is greater than $1$. It is evident that the
embedding predictor falls well short of being optimal, while the Padé
predictor approaches optimality.

\begin{figure}
\centering{}\includegraphics[scale=0.6]{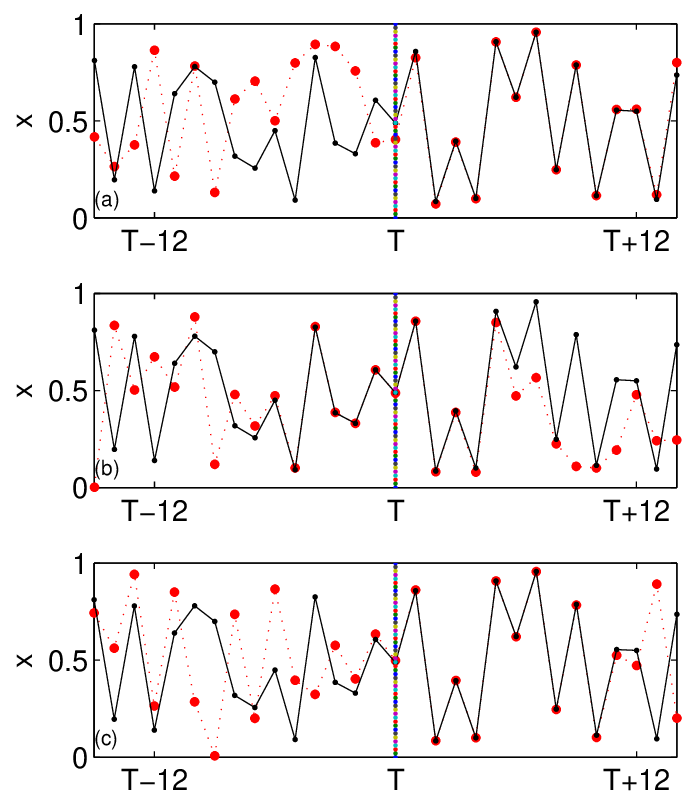}\caption{In each of the three plots, the black dots are part of a signal obtained
from the iterates of the automorphism \prettyref{eq:toral-auto-2d}
of the two dimensional torus $\mathbb{T}^{2}$. Here $T=2^{21}$.
The bigger red dots are: (a) the best fit from the past; (b) suboptimal
prediction using the embedding method; (c) optimal prediction using
the Padé method. \label{fig:torus-2d}}
 
\end{figure}

From Figure \ref{fig:torus-2d}, we see that the best fit from the
past does not agree too well with the signal at $T-1$, $T-2$, and
so on. However, it rapidly converges to the signal starting at $T$
and closely tracks the signal for more than $12$ iterates. The embedding
predictor on the other hand does too good a job of fitting the past,
but tracks only $5$ iterates from $T$ onwards. The Padé predictor
produces a match that requires a few iterates in the past to be close
enough for exponential extrapolation. Except for that, it reproduces
the behavior of the best fit where the signal segment that is chosen
from the history of the signal converges rapidly to the signal at
$t=T$ and then tracks it for a number of iterates.

\begin{table}
\begin{centering}
\begin{tabular}{|c|c|c|c||c|c|c|c|}
\hline 
$\log_{2}T$ & $t_{best}$ & $t_{embed}$ & $t_{pade}$ & $\log_{2}T$ & $t_{best}$ & $t_{embed}$ & $t_{pade}$\tabularnewline
\hline 
\hline 
5 & 3 & 0 & 0 & 17 & 12 & 0 & 8\tabularnewline
\hline 
6 & 5 & 0 & 2 & 18 & 14 & 0 & 13\tabularnewline
\hline 
7 & 1 & 1 & 0 & 19 & 15 & 0 & 11\tabularnewline
\hline 
8 & 4 & 1 & 1 & 20 & 15 & 7 & 15\tabularnewline
\hline 
9 & 6 & 0 & 2 & 21 & 15 & 0 & 12\tabularnewline
\hline 
10 & 4 & 0 & 3 & 22 & 18 & 0 & 14\tabularnewline
\hline 
11 & 8 & 2 & 2 & 23 & 17 & 1 & 16\tabularnewline
\hline 
12 & 11 & 0 & 4 & 24 & 18 & 9 & 18\tabularnewline
\hline 
13 & 10 & 2 & 4 & 25 & 20 & 8 & 19\tabularnewline
\hline 
14 & 8 & 0 & 6 & 26 & 20 & 6 & 18\tabularnewline
\hline 
15 & 11 & 2 & 8 & 27 & 21 & 6 & 19\tabularnewline
\hline 
16 & 13 & 6 & 9 & 28 & 23 & 7 & 21\tabularnewline
\hline 
\end{tabular}
\par\end{centering}

\caption{Length of best fit from the past, suboptimal prediction using the
method of embedding, and optimal Padé prediction of a signal obtained
from the automorphism \prettyref{eq:toral-auto-3d} of the three dimensional
torus $\mathbb{T}^{3}$.\label{tab:torus-3d}}
\end{table}
\begin{figure}
\centering{}\includegraphics[scale=0.6]{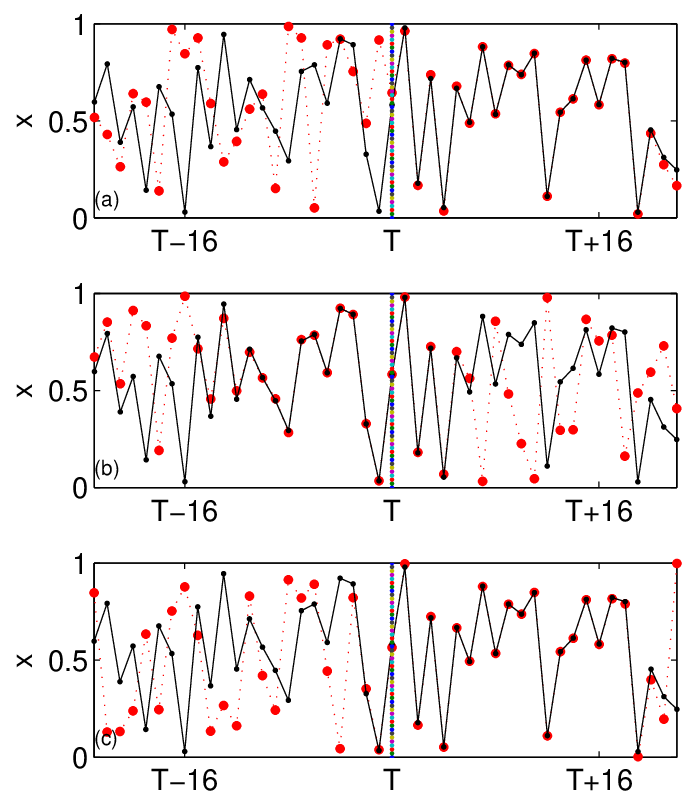}\caption{In each of the three plots, the black dots are part of a signal obtained
from the iterates of the automorphism \prettyref{eq:toral-auto-3d}
of the three dimensional torus $\mathbb{T}^{3}$. Here $T=2^{28}$.
The bigger red dots are: (a) the best fit from the past; (b) suboptimal
prediction using the embedding method; (c) optimal prediction using
the Padé method. \label{fig:torus-3d}}
\end{figure}

Table \ref{tab:torus-3d} and Figure \ref{fig:torus-3d} refer to
the toral automorphism defined by \prettyref{eq:toral-auto-3d}. By
going down Table \ref{tab:torus-3d} and comparing it with Table \ref{tab:torus-2d},
we notice that the automorphism of $\mathbb{T}^{3}$ has lower entropy
than the automorphism of $\mathbb{T}^{2}$. The tendency of the embedding
predictor to fit into the past is very pronounced in the middle plot
of Figure \ref{fig:torus-3d}.

The figures and tables of this section give a good sense of how much
is lost when a predictor fails to account for the unstable components
of the distance between segments of the signal. They also suggest
that a predictor which subjects the signal to more delicate analysis
should be able to approach optimality.

\section{Conclusion}

Matching the pattern of events leading to the present moment is a
natural idea for predicting nonlinear signals. Bode and Shannon \cite[1950]{BodeShannon1950}
expressed that idea as follows:
\begin{quotation}
The fact that nonlinear effects may be important in a prediction can
be illustrated by returning to the problem of forecasting tomorrow's
weather. We are all familiar with the fact that the pattern of events
over a period of time may be more important than the happenings taken
individually in determining what will come. For example, the sequence
of events in the passage of a cold or warm front is characteristic.
Moreover, the significance of a given happening may depend largely
upon the intensity with which it occurs. Thus, a sharp dip in the
barometer may mean that moderately unpleasant weather is coming. Twice
as great a drop in the same time, on the other hand, may not indicate
that the weather will be merely twice as unpleasant; it may indicate
a hurricane.
\end{quotation}
The central point of this paper is that a good predictor of chaotic
signals must not simply try to find a pattern of events that is as
close as possible to the pattern of events leading up to the current
time. The distance between the two patterns of events must be resolved
into stable and unstable components. The magnitudes of the unstable
components must be small and delicately balanced for optimal prediction.
The stable components on the other hand are typically as large as
the tolerance for correct prediction permits.

This conclusion has a counter-intuitive consequence. Because the stable
components are typically not small, the known pattern of events which
is best suited for predicting the current pattern of events will not
resemble the current pattern particularly closely.

\section{Acknowledgments}

The authors thank Emery Brown, John Gibson, Jeff Humphreys, Charles
Li, Steve Lalley, Roddam Narasimha and the referees for useful discussions.
We acknowledge support from NSF grants DMS-0715510, DMS-1115277, and
SCREMS-1026317.

\bibliographystyle{plain}
\bibliography{prediction}

\end{document}